\documentclass[sn-nature]{sn-jnl}% referee option is meant for double line spacing
\usepackage{multirow}%
\usepackage{amsmath,amssymb,amsfonts}%
\usepackage{amsthm}%
\usepackage{hyperref} 
\usepackage{manyfoot}%
\usepackage{mathrsfs}%
\usepackage{xcolor}%

\usepackage{txfonts}
\def\kms{km\,s$^{-1}$}

\def\zsun{Z$_{\odot}$}
\def\Zsun{Z$_{\odot}$}

\def\msun{M$_{\odot}$}

\def\l{$\lambda$}

\def\bloem{BLOeM}

\def\aap{Astron.~Astrophys.}

\def\aapr{Astron.~Astrophys.~Rev.}

\def\apj{Astrophys.~J}
\def\apjs{Astrophys.~J.~Suppl.}
\def\apjl{Astrophys.~Lett.}

\def\araa{Ann.~Rev. Astron.~Astroph.}
\def\mnras{Mon.~Not.~Royal~Ac.~Soc.}
\def\nat{Nature}

\def\pasa{Publ. Astron. Soc. Australia}

\begin{document}

\title{A high fraction of close massive binary stars at low metallicity}


\author[1,2,53]{H.\ Sana}
\author[3,53]{T.\ Shenar}
\author[4,5,53]{J.\ Bodensteiner} 
\author[6]{N.\ Britavskiy}
\author[7,8]{N.\ Langer}
\author[9,10]{D.\ J.\ Lennon}
\author[6]{L.\ Mahy}
\author[11,12]{I.\ Mandel}
\author[13]{S.\ E.\ de Mink}
\author[14]{L.\ R.\ Patrick}
\author[15]{J.\ I.\ Villase\~{n}or}
\author[1]{M.\ Dirickx}
%                 % alphabetical from here down
\author[9, 10]{M.\ Abdul-Masih}
\author[16]{L.\ A.\ Almeida}
\author[1]{F. Backs} 
\author[9,10]{S.\ R.\ Berlanas}
\author[17]{M.\ Bernini-Peron}
\author[18,1]{D.\ M.\ Bowman}
\author[19,20]{V.\ A.\ Bronner}
\author[21]{P.\ A.\ Crowther} 
\author[1]{K.\ Deshmukh}
\author[22]{C.\ J.\ Evans }
\author[1,23]{M.\ Fabry}
\author[24,25,26]{M.\ Gieles}
\author[27]{A.\ Gilkis }
\author[17]{G.\ Gonz\'alez-Tor\`a}
\author[7]{G.\ Gr{\"a}fener}
\author[28]{Y.\ G{\"o}tberg}
\author[29]{C.\ Hawcroft }
\author[30]{V.\ H\'enault-Brunet }
\author[9, 10]{A.\ Herrero}
\author[9, 10]{G.\ Holgado}
\author[31]{R.\ G.\ Izzard} 
\author[5, 1]{A.\ de Koter} 
\author[1,32]{S.\ Janssens}
\author[13,1]{C.\ Johnston}
\author[17]{J.\ Josiek}
\author[13]{S.\ Justham}
\author[33]{V.\ M.\ Kalari}
%\author[3}]{Z.\ Z.\ Katabi}
%\author[\ref{inst:naoj}]{Z.\ Keszthelyi}
\author[4]{J.\ Klencki} 
\author[34]{J.\ Kub\'at}
\author[34]{B.\ Kub\'atov\'a}
\author[17]{R.\ R.\ Lefever}
\author[35]{J.\ Th. van Loon}
\author[36, 1]{B.\ Ludwig}
\author[37]{J.\ Mackey}
\author[38]{J.\ Ma{\'\i}z Apell\'aniz} 
\author[39,40]{G.\ Maravelias} 
\author[1,41]{P.\ Marchant} 
\author[3]{T.\ Mazeh}
\author[42]{A.\ Menon}  
\author[43]{M.\ Moe}
\author[14]{F.\ Najarro}
\author[44]{L.\ M.\ Oskinova}
\author[3]{R.\ Ovadia} 
\author[44]{D.\ Pauli}
\author[45]{M.\ Pawlak}
\author[17]{V.\ Ramachandran}
\author[46]{M.\ Renzo}
\author[47]{D.\ F. Rocha} 
\author[17]{A.\ A.\ C.\ Sander}
%\author[x]{T.\ Sayada\inst{3}} 
\author[19,17]{F.\ R.\ N.\ Schneider}
\author[7]{A.\ Schootemeijer}
\author[17]{E.\ C.\ Sch\"osser}
\author[7]{C.\ Sch\"urmann}
\author[46]{K.\ Sen}
\author[48]{S.\ Shahaf}
\author[9, 10]{S.\ Sim\'on-D\'iaz} 
\author[49,50]{L.\ A.\ C.\ van Son}
\author[5]{M.\ Stoop}
\author[5]{S.\ Toonen}
\author[14]{F.\ Tramper}
\author[13]{R.\ Valli}
\author[13]{A.\ Vigna-G\'omez}       
\author[51]{J.\ S. Vink} 
\author[13, 52]{C.\ Wang}
\author[1]{R.\ Willcox}

%1
\affil[1]{Institute of Astronomy, KU Leuven, Celestijnlaan 200D, 3001 Leuven, Belgium} 
%2
\affil[2]{Leuven Gravity Institute, KU Leuven, Celestijnenlaan 200D, box 2415, 3001 Leuven, Belgium}
%3
\affil[3]{School of Physics and Astronomy, Tel Aviv University, Tel Aviv 6997801, Israel}
%4
\affil[4]{ESO - European Southern Observatory, Karl-Schwarzschild-Strasse 2, 85748 Garching bei M\"unchen, Germany}
%5
\affil[5] {Anton Pannekoek Institute for Astronomy, University of Amsterdam, Science Park 904, 1098 XH Amsterdam, the Netherlands}
%6
\affil[6]{Royal Observatory of Belgium, Avenue Circulaire/Ringlaan 3, B-1180 Brussels, Belgium}
%7
\affil[7]{Argelander-Institut f\"{u}r Astronomie, Universit\"{a}t Bonn, Auf dem H\"{u}gel 71, 53121 Bonn, Germany}
%8
\affil[8]{Max-Planck-Institut für Radioastronomie, Auf dem H\"ugel 69, 53121, Bonn, Germany}
%9
\affil[9]{Instituto de Astrof\'isica de Canarias, C. V\'ia L\'actea, s/n, 38205 La Laguna, Santa Cruz de Tenerife, Spain\label{inst:iac}}
%10
\affil[10]{Universidad de La Laguna, Dpto. Astrof\'isica, Av.\ Astrof\'sico Francisco S\'anchez, 38206 La Laguna, Santa Cruz de Tenerife, Spain}
%11
\affil[11]{School of Physics and Astronomy, Monash University, Clayton VIC 3800, Australia}
%12
\affil[12]{ARC Centre of Excellence for Gravitational-wave Discovery (OzGrav), Melbourne, Australia}
%13
\affil[13]{Max-Planck-Institute for Astrophysics, Karl-Schwarzschild-Strasse 1, 85748 Garching, Germany}
%14
\affil[14]{Centro de Astrobiolog\'ia (CSIC-INTA), Ctra.\ Torrej\'on a Ajalvir km 4, 28850 Torrej\'on de Ardoz, Spain}
%15
\affil[15]{Max-Planck-Institut f\"{u}r Astronomie, K\"{o}nigstuhl 17, D-69117 Heidelberg, Germany}
%16
\affil[16]{Escola de Ci{\^e}ncias e Tecnologia, Universidade Federal do Rio Grande do Norte, Natal, RN 59072-970, Brazil}
%17
\affil[17]{Zentrum f\"ur Astronomie der Universit\"at Heidelberg, Astronomisches Rechen-Institut, M\"onchhofstr. 12-14, 69120 Heidelberg, Germany} 
%18
\affil[18] {School of Mathematics, Statistics and Physics, Newcastle University, Newcastle upon Tyne, NE1 7RU, UK}
%19
\affil[19]{Heidelberger Institut f{\"u}r Theoretische Studien, Schloss-Wolfsbrunnenweg 35, 69118 Heidelberg, Germany}
%20
\affil[20]{Universit\"{a}t Heidelberg, Department of Physics and Astronomy, Im Neuenheimer Feld 226, 69120 Heidelberg, Germany}
%21
\affil[21]{Department of Physics \& Astronomy, Hounsfield Road, University of Sheffield, Sheffield, S3 7RH, United Kingdom}
%22
\affil[22]{European Space Agency (ESA), ESA Office, Space Telescope Science Institute, 3700 San Martin Drive, Baltimore, MD 21218, USA}
%23
\affil[23]{Department of Astrophysics and Planetary
Science, Villanova, PA, USA}
\affil[24]{ICREA, Pg. Llu\'{i}s Companys 23, E08010 Barcelona, Spain\label{inst:icrea}}
%24
\affil[25] {Institut de Ci\`{e}ncies del Cosmos (ICCUB), Universitat de Barcelona (IEEC-UB), Mart\'{i} i Franqu\`{e}s 1, E08028 Barcelona, Spain\label{inst:iccub}}
%25
\affil[26] {Institut d’Estudis Espacials de Catalunya (IEEC), Edifici RDIT, Barcelona, Spain\label{inst:ieec}}
\affil[27] {Institute of Astronomy, University of Cambridge, Madingley Road, Cambridge CB3 0HA, United Kingdom\label{inst:cambridge}}
%26
\affil[28]{{Institute of Science and Technology Austria (ISTA), Am Campus 1, 3400 Klosterneuburg, Austria}\label{inst:ista}}
%27
\affil[29] {Space Telescope Science Institute, 3700 San Martin Drive, Baltimore, MD 21218, USA\label{inst:stsci}}
%28
\affil[30] {Department of Astronomy and Physics, Saint Mary's University,    923 Robie Street, Halifax, B3H 3C3, Canada\label{inst:smu}}
%29
\affil[31] {Astrophysics Research Group, University of Surrey, Guildford, Surrey GU2 7XH, UK\label{inst:surrey}}
%30
\affil[32] {Research Center for the Early Universe, Graduate School of Science, University of Tokyo, Bunkyo, Tokyo 113-0033, Japan\label{inst:tokyo}}
%31
\affil[33] {Gemini Observatory/NSF's NOIRLab, Casilla 603, La Serena, Chile\label{inst:gemini}}
%32
\affil[34]{Astronomical Institute, Academy of Sciences of the Czech Republic, Fri\v{c}ova 298, CZ-251 65 Ond\v{r}ejov, Czech Republic\label{inst:ondrejov}}
%33
\affil[35]{Lennard-Jones Laboratories, Keele University, ST5 5BG, UK\label{inst:keele}}
%\affil[x]{Center for Computational Astrophysics, Division of Science, National Astronomical Observatory of Japan, 2-21-1, Osawa, Mitaka, Tokyo 181-8588, Japan\label{inst:naoj}}
%34
\affil[36]{Department of Astronomy and Astrophysics, University of Toronto, 50 St. George Street, Toronto, Ontario, M5S 3H4, Canada\label{inst:utoronto}}
%35
\affil[37]{Dublin Institute for Advanced Studies, DIAS Dunsink Observatory, Dunsink Lane, Dublin 15, Ireland\label{inst:dias}}
%36
\affil[38]{Centro de Astrobiolog\'ia (CSIC-INTA), ESAC campus, Camino bajo del castillo s/n, 28\,692 Villanueva de la Ca\~nada, Spain\label{inst:cabesac}}     
%37
\affil[39]{IAASARS, National Observatory of Athens, GR-15236, Penteli, Greece\label{inst:noa}}
%38
\affil[40]{Institute of Astrophysics, FORTH, GR-71110, Heraklion, Greece\label{inst:forth}}
%39
\affil[41]{Sterrenkundig Observatorium, Universiteit Gent, Gent, Belgium\label{inst:ghent}}
\affil[42]{Department of Astronomy, Columbia University, New York, NY, USA\label{inst:columbia}}
\affil[43]{University of Wyoming, Physics \& Astronomy Department, 1000~E.~University~Ave., Laramie, WY 82071, USA\label{inst:wyoming}}
%40
\affil[44]{Institut f\"ur Physik und Astronomie, Universit\"at Potsdam, Karl-Liebknecht-Str. 24/25, 14476 Potsdam, Germany\label{inst:up}}
%41
\affil[45]{Lund Observatory, Division of Astrophysics, Department of Physics, Lund University, Box 43, SE-221 00, Lund, Sweden\label{inst:lund}}
%42
\affil[46]{Department of Astronomy \& Steward Observatory, 933 N. Cherry Ave., Tucson, AZ 85721, USA\label{inst:AZ}}
%43
\affil[47] {Observat\'orio Nacional, R. Gen. Jos\'e Cristino, 77 - Vasco da Gama, Rio de Janeiro - RJ, 20921-400, Brazil\label{inst:ON_Br}}
%44
%\affil[\ref{inst:umk}]{{Institute of Astronomy, Faculty of Physics, Astronomy and Informatics, Nicolaus Copernicus University, Grudziadzka 5, 87-100 Torun, Poland}\label{inst:umk}}
%45
\affil[48]{Department of Particle Physics and Astrophysics, Weizmann Institute of Science, Rehovot 7610001, Israel\label{inst:weizmann}}
%46
\affil[49]{Center for Computational Astrophysics, Flatiron Institute, New York, NY 10010, USA\label{inst:cca}}
\affil[50]{Department of Astrophysical Sciences, Princeton University, 4 Ivy Lane, Princeton, NJ 08544, USA\label{inst:princeton}}
%47
\affil[51]{Armagh Observatory, College Hill, Armagh, BT61 9DG, Northern Ireland, UK\label{inst:armagh}}
%4
\affil[52]{Department of Astronomy, Nanjing University, Nanjing 210023, People’s Republic of China\label{inst:nanjing}}
\affil[53]{These authors contributed equally to this work. Correspondence should be addressed to H. Sana (\href{hugues.sana@kuleuven.be}{hugues.sana@kuleuven.be}), T. Shenar (\href{tshenar@tauex.tau.ac.il}{tshenar@tauex.tau.ac.il}) and J. Bodensteiner (\href{j.bodensteiner@uva.nl}{j.bodensteiner@uva.nl})}

%%%%%%%%%%%%%%%%%% ABSTRACT %%%%%%%%%%%%%%%%%%
\abstract{ 
%The interpretation of observations of gravitational waves, transients, and distant galaxies in the high-redshift Universe heavily relies on our understanding of massive stars at low metallicity. 
At high metallicity, a majority of massive stars have at least one close stellar companion. %\cite{KK2014,own,RCN2022,MB2024,OMH2023}. 
The evolution of such binaries is subject to strong interaction processes, heavily impacting the characteristics of their life-ending supernova and compact remnants. %\cite{Pols1994,Podsiadlowski1992,Eldridge2022ARA&A}. 
For the low-metallicity environments of high-redshift galaxies constraints on the multiplicity properties of massive stars over the separation range leading to binary interaction are crucially missing.  Here we show that the presence of massive stars in close binaries is ubiquitous, even at low metallicity. 
%The Small Magellanic Cloud is a  low-metallicity dwarf galaxy neighbouring our Milky Way with a metal content of about one fifth of the solar value, \cite{hunter2007}, hence similar to that of massive galaxies before and during the peak of star formation in the Universe \cite{LN2006,Madau2014,Stanton2024}. 
Using the Very Large Telescope, we obtained multi-epoch radial velocity measurements of a representative sample of 139 massive O-type stars across the Small Magellanic Cloud, which has a metal content of about one fifth of the solar value. We find that 45\%\ of them show radial velocity variations which demonstrate that they are members of close binary systems, and predominantly have orbital periods shorter than one year. Correcting for observational biases indicates
that at least $70^{+11}_{-6}$\%\ of the O stars in our sample are in close binaries, and that at least $68^{+7}_{-8}$\%\ of all O stars interact with a companion star during their lifetime. We found no evidence supporting a statistically significant trend of the multiplicity properties with metallicity.
Our results indicate that multiplicity and binary interactions govern the evolution of massive stars and determine their cosmic feedback and explosive fates. %This has strong implications for our understanding of the high redshift Universe, and of the incidence rate and properties of observable gravitational wave signals.
}

%%%%%%%%%%%%%%%%%% END OF ABSTRACT %%%%%%%%%%%%%%%%%%

\maketitle
\section*{Main text}
%%%%%%%%%%%%%%%%%%%%%% Paragraph 1: Metallicity is key in astrophysics
Baryonic matter in the Universe just after the Big Bang consisted almost purely of hydrogen and helium, with only a tiny fraction of less than $10^{-9}$ in heavier
elements, so-called metals. The bulk of the heavier elements, and foremost oxygen, have since been produced in the interiors of massive stars (stars born with more than $\sim 8\,$M$_{\odot}$ \cite{Poelarends2008}),
and released into the interstellar medium in their explosive end stages \cite{Burbidge1957, Nomoto2013}. While low- and intermediate-mass stars also contribute to chemical evolution, massive stars, including the crucial contribution of merging neutron stars, are thought to be responsible for forming the majority of elements in the periodic table \cite{Kobayashi2020}. Moreover, the pace of the overall metallicity enrichment in the Universe is set by short-lived massive stars, which results in a strong correlation between the star formation history of the Universe and its metal content \cite{Kobayashi2007, Maiolino2019, Wilkins2023}. \\ 

Massive stars are important physical ingredients of our Universe, and hence of large-scale cosmological models. Their radiation heats and ionises the interstellar medium \cite{Hopkins2012}, and contributes to the re-ionisation of the Universe at redshifts beyond $z\simeq 5$ \cite{Bosman2022}. Photo-heating, radiation pressure and end-of-life explosions affect the formation and evolution of the first galaxies and regulates the efficiency of star formation in the turbulent interstellar medium \cite{Hopkins2014}. Supernovae and massive-star clusters are the main sources of cosmic rays in galaxies \cite{Vieu2023}, which strongly impact the dynamics of galactic disks and contribute to driving galactic-scale outflows \cite{Girichidis2018}. Massive stars also dominate the integrated light of star-forming galaxies \cite{starburst99}, produce neutron stars and black holes, as well as X-ray binaries \cite{Heger2003} and transient gravitational-wave sources \cite{gwtc3,Fortin2024}.\\

%%%%%%%%%%%%%%%%%%%%%% Paragraph 3 

In our Milky Way, the vast majority of massive stars are found in binaries or higher-order multiple systems \cite{sana2012,KK2014,sana2014,own,RCN2022,MB2024,OMH2023}. In massive stars in close binaries, that is binaries with an orbital period $P_\mathrm{orb} \lesssim 4$ years (see final discussion in Methods), the presence of a nearby companion has strong consequences for the way the two stars evolve: they will exchange mass and angular momentum, and some will even merge \cite{Pols1994,Podsiadlowski1992}. These interactions strongly affect the lifetime, radiative feedback and final fate of massive stars  \cite{Langer2012,Eldridge2022ARA&A}. They also modify the appearance of entire populations of massive stars seen in integrated light \cite{Eldridge2017,Gotberg2019}, and the types of supernovae produced \cite{Eldridge2008}. Establishing whether the profusion of massive close binaries in our Milky Way persist in low metallicity environments is fundamental for  understanding the early Universe, and for the accuracy of our cosmic formation and evolution models.\\

%%%%%%%%%%%%%%%%%%%%%% Paragraph 4 : LMC and SMC as a priviledge laboratories. 
The Large (LMC) and Small Magellanic Cloud (SMC) are two neighboring dwarf galaxies with a current metal content of about one-half (\zsun/2) and one-fifth Solar (\zsun/5), respectively \cite{hunter2007}.  At any redshift (a look-back time equivalent), massive star-forming galaxies are, on average, more metal-rich than low-mass galaxies \cite{2010MNRAS.408.2115M}. Indeed, the Milky Way, the LMC and the SMC follow this mass-metallicity relation \cite{AM2013, Curti2023}. The average metallicity in the Universe  further decreases with redshift. The LMC and SMC metallicities are thus not representative of the metal content in the local Universe, but rather correspond to the metallicity of massive star-forming galaxies at a redshift of 0.5...1 and of 3...10, respectively (e.g.\ \cite{LN2006, Morishita2024, Nakajima2023, Heintz2023, Li2023}, and references therein).
Only very rarely can stars be individually studied in such high-redshift galaxies \cite{Icarus2018, Earendel2022}.  In particular, and with current observational capabilities, the SMC is the only galaxy close enough to measure binary properties of representative stellar samples in a  low-metallicity environment.\\

Several efforts have investigated the metallicity dependence of binarity for low mass stars \cite{Badenes2018,Moe2019,Niu2022,Price-Whelan2020}, revealing an increase of the solar-type binary fraction from 0.1 to 0.4 as the metallicity decreases from 3\,\Zsun\, to 0.1\,\Zsun, hence an approximate slope of $-0.2/\mathrm{dex}$ \cite{Moe2019}.\\

Hydrodynamic simulations suggest that massive protostellar disks are likely to fragment into binary companions at SMC metallicity \cite{Tanaka2014}, but will preferentially form distant companions first. 
The processes of protobinary fragmentation, circumbinary accretion, and orbital migration are however quite uncertain \cite{OMH2023}, motivating the need for empirical measurements. The eclipsing binary fraction of unevolved early B-type stars (i.e., main-sequence massive stars with a typical birth mass in the range of 8 to 15~\msun) is consistently around 1 to 2\%\ across the Milky Way, LMC, and SMC \cite{Moe2013}, with no significant trend with metallicity. Detection of eclipsing binaries is, however, strongly limited  by geometrical effects to very short orbital periods ($P_\mathrm{orb} \lesssim 20$~days) and high orbital inclinations ($i \gtrsim 65^\mathrm{o}$), and does not allow to probe the full range of periods relevant to binary formation and evolution (see Methods). Like the Milky Way, the LMC also shows a large fraction of massive stars in binaries close enough to interact \cite{vfts8,vfts11,tmbm1}. While high, the LMC fraction is slightly smaller than that of the Milky Way,  raising the question of whether massive binaries remain frequent in low metallicity environments or whether the perceived downward trend between the Milky Way and LMC environments persists at lower metallicity.
\\

%%%%%%%%%%%%%%%%%%%%%% Paragraph 5: The BLOeM survey and the O star sample
In this context, we have used the Very Large Telescope of the European Southern Observatory (ESO) to obtain multi-epoch spectroscopy of over 900 massive stars spread across the SMC, including O and B main-sequence stars, and B, A and F supergiants. Gathered as part of the Binarity at LOw Metallicity (\bloem) large program (ESO program ID: 112.25R7, PIs: Shenar/Bodensteiner) \cite{bloem1}, the observations used the FLAMES/Giraffe spectrograph \cite{Pasquini2002}, a multi-fiber instrument allowing us to obtain optical spectra of over 100 stars simultaneously across a 20' field of view. The survey, data acquisition, and data reduction are described in a separate work \cite{bloem1}. More details are provided in the Methods section. Example spectra are provided on Extended Data Figure~1. %\ref{f:spectra}. 
\\

\begin{figure*}[h]
%\centering
\includegraphics[width=0.99\linewidth]{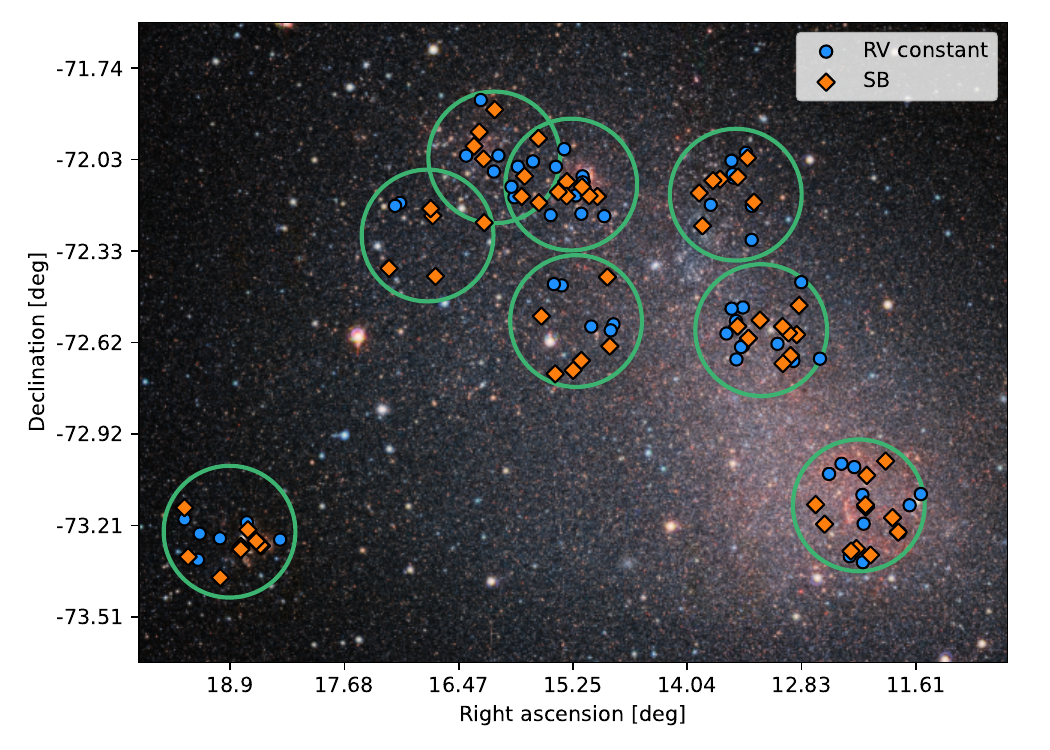}
\caption{\textbf{Distribution of O-type stars} in the \bloem\ sample overlaid on a VISTA Y-J-K$_\mathrm{S}$ false-color image of the SMC. Image credit:ESO/VISTA VMC. Large circles show the 8 fields of view of the \bloem\ campaign. Diamonds indicate detected O-type spectroscopic binaries (SB); circles, RV constant, presumably single stars. }
\label{f:fov}
\end{figure*}

Here, we focus on the 139  O-type stars observed at nine different epochs between October and December 2023 (Figure~\ref{f:fov}) \cite{bloem1}. The \bloem\ O-star sample does not suffer from incompleteness due to the brightness limit of the survey \cite{bloem1} and the low extinction towards the SMC. The higher luminosities and higher masses of O stars further provide higher signal-to-noise spectra and a clearer binary signal than main sequence B-type stars \cite{bloem_Bstars}. The O-type star sample is also less impacted by evolutionary effects than the more evolved B-A-F supergiants present in \bloem\ \cite{bloem_BAFsg} and is thus more representative of initial conditions.  Finally, focusing on the O stars allows to directly compare with similar studies at higher metallicity in the Milky Way (\zsun) and the LMC (\zsun/2). 
\\

%%%%%%%%%%%%%%%%%%%%%% Paragraph 6: RV measurements MC detection probability and the sensitivity to P < 1-2 years

We use cross-correlation and line-profile fitting to derive the radial velocities (RVs) at each epoch (see Methods). We then identify the stars that displayed significant RV variations ($\Delta RV > 4\sigma_{\Delta RV}$) with an amplitude of at least 20~\kms\ ($\Delta RV > 20 \rm{km\,s}^{-1}$). This threshold is large enough to avoid false positives due to measurement uncertainties and other possible sources of variability \cite{vfts8}, and it is identical to that used in other studies, allowing for a fair comparison of other binary detection measurements. Under these criteria, we identify 62 stars displaying significant, large RV variability, revealing that they are most likely part of a spectroscopic binary system. This corresponds to an observed binary fraction of $f_\mathrm{obs}=0.45\pm0.04$, where the $\pm1\sigma$ uncertainty is computed using binomial statistics and the sample size and corresponds to the 68\%-confidence interval. \\

\begin{figure*}[h]
\centering
\includegraphics[width=0.99\linewidth]{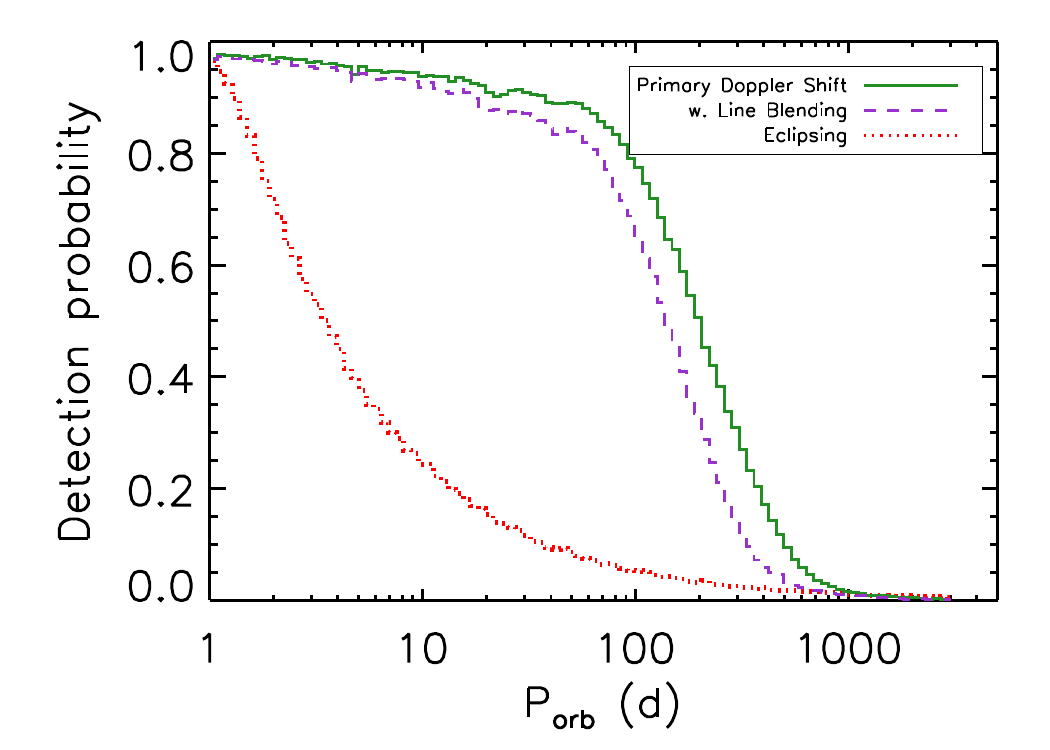}
\caption{{Binary detection probability} of the \bloem\ survey for O stars in our sample as a function of the orbital period. The plain green line is computed while considering the Doppler shift of the primary star only (main method). The dashed, purple line also include the line-blending detection bias, which reduces detection at long periods. The dotted, red lines indicates the fraction of the simulated systems that display eclipses according to Equation \ref{eq:eclipses}. Detection probability curves as function of other parameters are provided in Extended Data Figure~2.}%\ref{f:detect}.}
\label{f:detect_p}
\end{figure*}

We simulate the detection probability of the \bloem\ survey for different orbital properties (orbital period, mass ratio, eccentricity and primary mass) using a Monte Carlo population synthesis code \cite{vfts8} and applying the \bloem\ observational sampling and measurement uncertainties. We show that our study is mostly sensitive to binaries with an orbital period shorter than about one year, dropping significantly at longer orbital periods (Figure~\ref{f:detect_p}). At orbital periods shorter than 100~days, we are sensitive to secondaries with masses as small as $\sim 10\%$ of the primary O star mass (Extended Data Figures~2 and 3).%~\ref{f:detect_2d}). 
This implies that most of our detected binaries are short-period systems that inevitably interact before either companion undergoes core collapse.
\\

%%%%%%%%%%%%%%%%%%%%%% Paragraph 7 : The intrinsic binary fraction 

\begin{figure*}[h]
\centering
\includegraphics[width=0.99\linewidth]{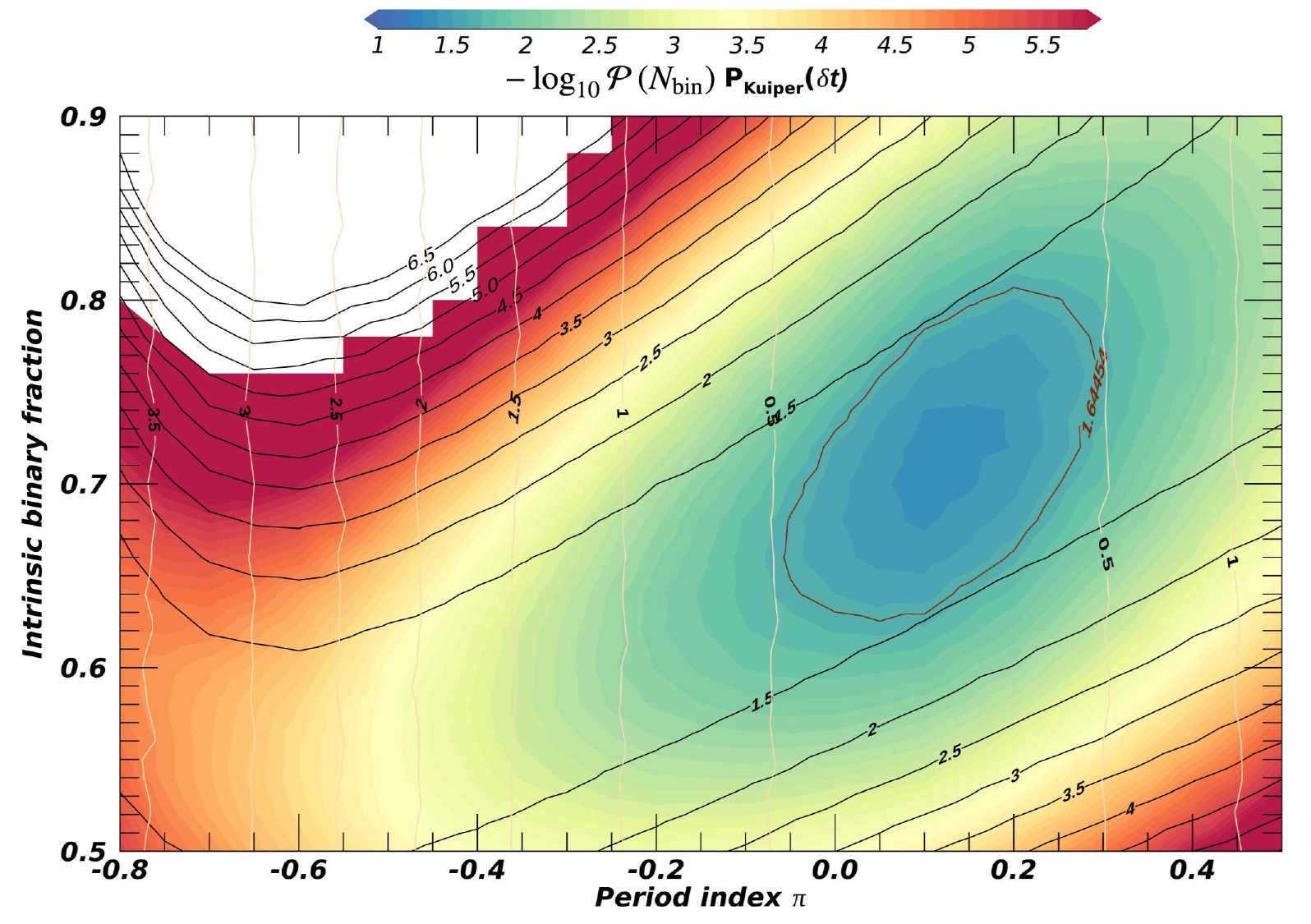}
\caption{2D map combining the binomial probability $\mathcal{P}$ and Kuiper statistics $P_\mathrm{Kuiper}$ of simulated populations to reproduce the observable multiplicity properties of our \bloem\ sample. The equi-probability contours of the two-sided Kuiper statistics (vertical lines) and the the probability to detect the same number of binaries as we do (diagonal lines) are overlaid. The color-scale give the products of both probabilities. The red contour indicates the half-peak value.
}
\label{f:proba_combined}
\end{figure*}

The present data set is insufficient to reliably measure the individual orbital properties of most of the binaries that we detect, however statistical constraints on the intrinsic binary fraction and orbital period distribution can be obtained. We synthesize various mock populations of O-stars, varying their intrinsic binary fraction and orbital period distribution,  applying the observational biases and binary detection criteria of the \bloem\ survey, and comparing the observational yields of these mock observing campaigns to actual data (see Methods). Here, we focus on reproducing the observed binary fraction ($f_\mathrm{bin}$) and the distribution of smallest time differences ($\delta t$) between observations that exhibit significant ($\Delta RV > 4\sigma_{\Delta RV}$) and large $(\Delta RV > 20 \rm{km\,s}^{-1})$  RV variations. The latter is an unbiased estimator of the shape of the period distribution function which has been validated using artificial data sets \cite{vfts8} (see also Extended Data Figure~4).%\ref{f:dHJD}). 
We vary the intrinsic binary fraction ($\mathcal{F}_\mathrm{bin}$) from 0.50 to 0.90 by step 0.02. We adopt a power-law function to describe the distribution of orbital periods ($f_\mathrm{\log P} \propto (\log P)^\pi$), and vary $\pi$ from $-0.80$ to $+0.50$ in steps of 0.05. We found that the populations with the best agreement with both the observed binary fraction (Extended Data Figure~5,%\ref{f:proba1}, 
upper panel) and the distribution of shortest time differences Extended Data Figure~5, %\ref{f:proba1}, 
lower panel) are characterized by an intrinsic binary fraction of $\mathcal{F}_\mathrm{bin}=0.70^{+0.11}_{-0.06}$ and an index of the orbital period distribution $\pi=+0.10^{+0.20}_{-0.15}$ (Figure~\ref{f:proba_combined}). Accounting for a line-blending detection bias affecting (near-)equal brightness binaries, which was not accounted for so far (see Methods),  would add another 5\%\ to the intrinsic binary fraction so that the value that we obtain is a lower limit.
\\

%%%%%%%%%%%%%%%%%%%%%% Paragraph 9 :  Comparison with higher-Z constraints; conclusion that it points towards universal
The methods that we applied to identify binaries and constrain the multiplicity property of the parent populations are similar to those used in higher metallicity studies. This allows for a direct and robust comparison of the outcome of these studies, providing clues to whether the metallicity of the environment is an important factor impacting massive-star multiplicity. The intrinsic O-type star binary fraction was derived for systems with orbital periods less than 8.7 years in young open clusters in the Milky Way (\zsun) \cite{sana2012}, obtaining $\mathcal{F}_\mathrm{bin}=0.69\pm0.09$.  Using the same instrumentation and methods that we apply here, the VLT-FLAMES Tarantula Survey (VFTS, \cite{vfts1}) measured the O-type star binary fraction in the 30~Doradus massive star-forming region (\zsun/2) in the LMC and obtained $0.51\pm0.04$ \cite{vfts8}. The latter value was revised to 0.58 after improving the orbital period distribution measurement \cite{tmbm1}. The Milky Way and LMC  intrinsic fraction of O-type  binaries are thus comparable to that of the SMC (\zsun/5) determined in this work. Similarly, the orbital period distributions are compatible, especially after accounting for minor differences in the minimum adopted period (see the discussion in Methods). \\

\begin{figure*}[h]
\centering
\includegraphics[width=\linewidth]{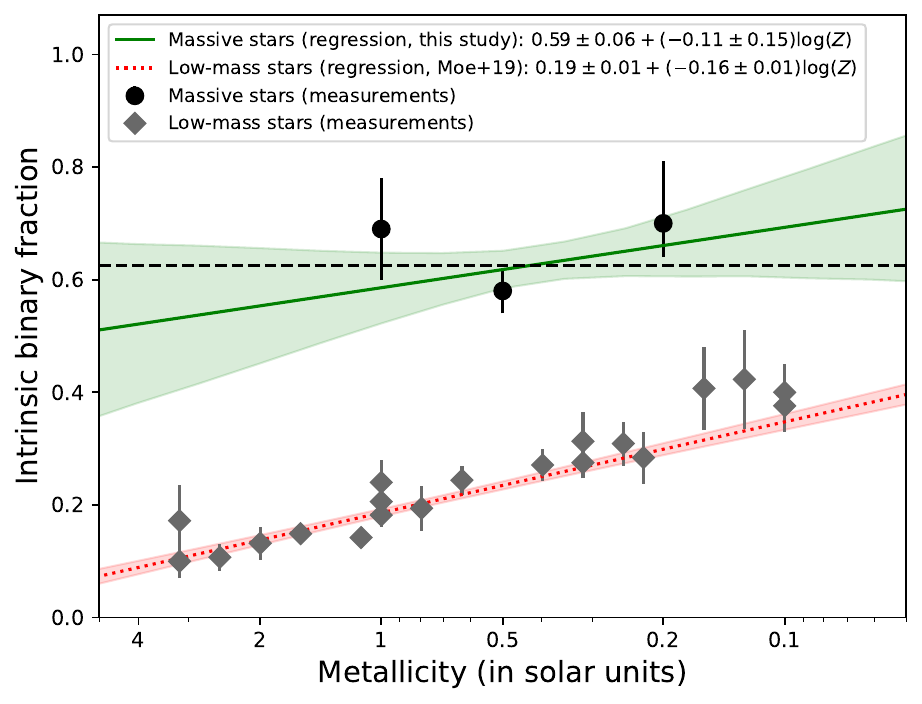}
\caption{Metallicity dependence of the intrinsic binary fraction of O-type stars (data: circles, model: green; this work) and of solar-mass stars (data: diamonds, model: red; \cite{Moe2019}). 
The best-fit parameters and their standard deviations are computed with a linear regression to the displayed data points and their 1$\sigma$ error-bars. The green and red shaded regions show the 68\%\ (1$\sigma$) confidence intervals of the regressions to the high-mass and low-mass data, respectively. The slope for massive stars ($-0.11 \pm 0.15$, solid line) is consistent with a constant line (dashed line), but does not allow to reject a slight trend with metallicity at the level of the slope reported for solar-mass stars ($-0.16 \pm0.01$, dotted line).   }
\label{f:reg}
\end{figure*}

To quantitatively assess a possible trend of the intrinsic fraction of close O-type binaries with metallicity, we performed a linear regression $\mathcal{F}_\mathrm{bin}=a+b\log_{10} (Z/Z_\odot)$ using the available Milky Way and LMC bias-corrected measurements \cite{sana2012,tmbm1} together with our new results (see Methods). We obtain no significant metallicity dependence (regression slope $b=(-0.11\pm0.15)/\mathrm{dex}$; see Figure~\ref{f:reg}). Albeit we cannot reject the slope of $(-0.16\pm0.01)/\mathrm{dex}$ measured from solar-mass stars data \cite{Moe2019} either, our results firmly show that the large abundance of close binaries` observed among massive stars in high-metallicity environments is also present at lower metallicity. \\

While the formation mechanism leading to a large fraction of massive close binaries remains an open question \cite{OMH2023}, our study  indicates that this process is unlikely to strongly correlate with metallicity. This yields empirical support for promising recent theoretical  efforts investigating the (lack of) metallicity dependence of star formation in general  (e.g., \cite{Hennebelle2024}) or more specifically of massive binaries (e.g., \cite{Guszejnov2023,Chon2024}). However, these simulations are so far only able to probe wider systems, emphasizing the need for empirical constraints at close separation. Constraints on close binaries are also crucial to decide whether the components of the binaries do interact or rather evolve as isolated single stars. Together with the existing multiplicity studies of different Milky Way and LMC environments, our results indicate that massive binaries are ubiquitous, yet that small differences may occur as a function of the star formation conditions (i.e., starburst vs.\ field) as suggested by the smaller close binary fraction measured in the Tarantula massive star forming region of the LMC \cite{vfts8,vfts11} compared to lower-mass open clusters in the Milky Way \cite{sana2012} and in this work (see Figure~\ref{f:reg}).
\\

%%%%%%%%%%%%%%%%%%%%%% Paragraph 10 : Impact of the results
We estimate the fraction of massive stars born as O stars that will interact with a nearby companion, either as mass donor, mass gainer, or through coalescence. This fraction is readily obtained by integrating the orbital parameter distributions and adopting our inferred intrinsic binary fraction (see Methods). We adopt a uniform mass ratio distribution, consistent with previous studies \cite{sana2012,tmbm6}, and we conservatively assumed that all systems with an orbital period less than 1500 days interact. Our results indicate that $68^{+7}_{-8}$\%\ of all SMC stars born as O stars will interact with their companion,  18\%\ of which ($12\pm4$\%\ of all O stars) will do so before leaving the main sequence.  Given the conservative assumptions made (see discussion in Methods), the percentage that we compute likely represents a lower limit. Our results focus on the global number of massive stars that interact and call for comparative evolutionary computations to assess the boundaries of specific binary evolutionary channels in various metallicity environments.
\\

%%%%%%%%%%%%%%%%%%%%%% Paragraph 11: Concluding paragraph
Our results imply that massive star evolution in the Universe is likely dominated by close binary star interactions out to high redshifts. This has strong consequences for stellar wind-induced feedback processes in galaxies  as well as the distribution and properties of single and binary neutron stars and black holes. The dominance of massive binaries at low-metallicity  increases the viability of binary evolution channels to the formation of gravitational wave events. Mass gainers and mergers lead to more massive stars that can produce additional feedback, while envelope stripping by mass transfer in massive binaries will occur abundantly in the high-redshift Universe. This leads to hot stripped stars \cite{Gotberg2020} and X-ray binaries \cite{Jeon2014,Sartorio2023}, which produce photons available for the re-ionization of the Universe \cite{Stanway2016}. 

\clearpage

%%%%%%%%%%%%%%%%%%%%%%%%%%%%%%%%%%%%%%%%%%%%%%%%%%%%%%%%%%%%
\section*{Methods}
\label{sec:method}

\subsection*{Sample and RV measurements}\label{sec:sample}
The \bloem\ survey targets over 900 massive star candidates in the SMC, based on magnitude and color cuts in a Gaia DR3 color-magnitude diagram \cite{bloem1}. The densest star-formation regions were avoided due to instrumental constraints so that \bloem\ targets are predominantly found in lower-density, high-mass star-forming regions of the SMC. The \bloem\ spectra were obtained with the LR02 setup of the VLT FLAMES/Giraffe spectrograph and cover the spectral range from 395 to 455~nm with a spectral resolving power $R=\lambda/\delta \lambda=6200$. This spectral range contains numerous diagnostic lines of hydrogen and helium, so is well suitable for RV determination. For each target we use the individual epoch spectra obtained by co-adding two 10-minute back-to-back exposures. 
Calibration, data reduction and spectral classification are described in \cite{bloem1}. In this work, we focus on the sample of 139 SMC targets classified as O-type stars. Each object has been observed at nine epochs between October and December 2023, providing an excellent temporal sampling to study variability on time scales of days, weeks and months.  

Most RV measurements were performed using a cross-correlation method \cite{zucker2003}, which was already successfully applied to O-type stars in, e.g., \cite{tmbm6}. In brief, for a given object, we first cross-correlate all spectra with that of the highest S/N epoch. After this first step, we shift and add all spectra to create a master spectrum with an even higher S/N. The latter is then used as new cross-correlation template for a second iteration of the cross-correlation process. More details are given in \cite{tmbm6}. The median 1$\sigma$ RV uncertainty is 2.5~\kms\ while 80\%\ (resp.\ 95\%) of the RV measurements have a 1$\sigma$ uncertainty below 5~\kms\ (resp.\ below 9~\kms). 

In 22 cases, the double-lined spectroscopic binary nature (SB2) of the objects was identified through visual inspection. In these cases, we used line profile fitting following the same general approach as in \cite{vfts8}. Specifically, each stellar line is fitted with the same Gaussian profile at all epochs while all lines of a given epoch are required to yield the same RV shift.  The primary was assumed to be the star displaying the relative RV shifts with the lowest amplitude.  This method is more robust than individual line-by-line and epoch-per-epoch line profile fitting but  not as robust as spectral disentangling \cite{tmbm6}. However, simultaneous line-profile fitting at all epochs requires no apriori knowledge of the orbital solution, which would be difficult to obtain for a large fraction of the current sample. 

We evaluate the precision of the error estimate by simulating artificial spectra with representative noise and applying the   cross-correlation template. We find that, in cases of well-behaved Gaussian noise, the dispersion of the measurements is within 10\%\ of the uncertainty estimates from \cite{zucker2003}'s method. Similar simulations were already performed in the literature for the cross-correlation \cite{zucker2003} and line profile fitting \cite{vfts8} approaches, with similar results.

The journal of the observations, the RV measurements and their uncertainties is provided in Table 1 which is available electronically at the Centre de Donn\'ees astrophysiques de Strasbourg (CDS; {\tt http://cdsweb.u-strasbg.fr}). Example spectra are displayed in Extended Data Figure~1.%\ref{f:spectra}. 

\subsection*{Observed binary fraction}
We follow the strategy outlined in \cite{vfts8} to detect spectroscopic binaries based on their RV variability. Specifically, we compute the significance of RV variability by comparing each pair of RV measurements and flag a star as RV variable if the RV variation for at least one of the 36 combinations of $i$ and $j$ corresponding to the nine different epochs is larger than 4$\sigma$,
\begin{equation}\label{eq:RVsignif}
\frac{|v_i-v_j|}{\sqrt{\sigma_i^2+\sigma_j^2}} > 4\,,
\end{equation}
where $i$ and  $j$ represent individual epochs of measurements, and $v$ and $\sigma$ are the RVs and the measurement uncertainties at the corresponding epochs.

Multiple physical processes can result in a RV variability signal, including instrumental artefacts, winds, pulsations and other sources of line-profile variability. We therefore applied a minimum threshold $C$ to the amplitude of significant RV variations and considered as reliable binary candidates only stars for which at least one pair of RV measurements ($v_i,v_j$) satisfies simultaneously the significance criteria of Eq.~\ref{eq:RVsignif} and $|v_i-v_j| > C$, where we adopt $C=20$~\kms\ for consistency with earlier studies \cite{sana2012,vfts8}. 79~systems show significant RV variations, 17~of them, however, do not meet our minimum RV amplitude criterion. Their RV variability amplitude is too small to be confidently assigned to binarity and some of this low-RV amplitude sample are possibly false binary detections. In the following, we focus on the 62~objects showing high-amplitude RV variability and we use bias-correction methods to recover the part of the binary population that presents low-amplitude RV variability. %This yields an observed binary fraction of f_\mathrm{obs}0.45\pm0.04$ for O-type stars in the \bloem\ SMC sample, where binomial statistics has been used to estimate the statistical uncertainties due to the sample size. 

Table 2, available at CDS, provides the list of RV variable and spectroscopic binaries that we identified. The detected binaries are well spread across the eight fields of view investigated by \bloem\ (see Figure~\ref{f:fov}), with detected fractions in each field ranging from $0.33\pm0.11$ to $0.50\pm0.15$. Given the small sample sizes in each field, we detect no statistically significant differences in the binary fraction measured per field and thus only consider the full sample in the rest of this work.

Pre-empting the discussion of observational biases provided below, the \bloem\ observed binary fraction for O stars, $f_\mathrm{obs}=0.45\pm0.04$, is in the same range as other O-star spectroscopic campaigns in the Galaxy (e.g., young Galactic clusters $0.56\pm0.06$ \cite{sana2012}, the OWN survey: $0.50\pm0.03$ \cite{own}, Cyg. OB2: $0.51\pm0.07$ \cite{KK2014}) or the LMC (VFTS survey: $0.35\pm0.03$ \cite{vfts8}); see overviews in, e.g., \cite{Sana2017, MB2024}.

\subsection*{Survey detection capability}
We used the Monte Carlo population-synthesis method presented in \cite{vfts8} to quantify the ability of the \bloem\ campaign to detect O-type binaries of different orbital periods , mass ratios  and eccentricities. We specifically adopted an improved version of the method that includes uncertainties on the adopted underlying distributions \cite{Banyard2022}. Specifically, we simulated 10\,000 observing campaigns of 139 O-type stars, adopting the temporal sampling and RV uncertainties from the \bloem\ data.  We used power-law representations for the distributions of orbital period $P_\mathrm{orb}$, mass ratio $q=M_2/M_1$, eccentricity $e$, and primary mass $M_1$, parameterized as follows:
\begin{eqnarray}
    f_{\log P} &\propto& \left(\log_{10} P\right)^\pi,\ \mathrm{with}\ \log_{10} \left(P\mathrm{/d}\right)=0.0...3.5\ \mathrm{and}\ \pi=+0.1\pm0.2, \label{eq:Pidx}\\
    f_{q} &\propto& q^\kappa,\ \mathrm{with}\ q=0.1...1.0\ \mathrm{and}\ \kappa=0.0\pm0.2, \label{eq:Qidx}\\
    f_{e} &\propto& e^\eta,\ \mathrm{with}\ e=0.0...0.9\ \mathrm{and}\ \eta=-0.5\pm0.2, \label{eq:Eidx}\\
     f_{M_1} &\propto& M_1^\gamma,\ \mathrm{with}\ M_1=15...60~\mathrm{M}_\odot\ \mathrm{and}\ \gamma=-2.35. \label{eq:Midx}
\end{eqnarray}
In each simulated campaign, we further varied the indexes of the underlying distributions of orbital parameters following normal distributions with 1$\sigma$ dispersions as specified in Eqs.~\ref{eq:Pidx}-\ref{eq:Eidx}. Circularization of the shortest-period systems is further accounted for following \cite{Banyard2022}.  Specifically, very short periods systems ($P_\mathrm{orb}<3$~days) are considered circularized ($e=0$), see e.g.\ \cite{sana2012,tmbm1,Lennon24}. Eccentricities that would lead to a periastron separation of less than 20 solar radii (20~R$_\odot$) are rejected and a new value is redrawn until this condition is met.

The index of the mas-ratio and eccentricity distributions $\kappa$ and $\eta$ cannot be constrained with the current data and we adopt values derived from Milky Way and LMC studies \cite{sana2012,tmbm1,tmbm6}. The choice of the period distribution index $\pi$ is discussed below. We included binaries with mass ratio $<0.1$ in simulations displayed on Extended Data Figures~2 and 3.%\ref{f:detect} and \ref{f:detect_2d}. 
In the following however, we exclude such extreme mass-ratios from the binary statistics as these systems are difficult to detect and their existence in close binaries remains debated.  Such low-mass companions are quickly swallowed upon interaction, presumably with limited evolutionary consequences hence we neglect them. Accounting for their presence only strengthens our conclusions.

To qualitatively compare the performances of spectroscopic surveys such as \bloem\ to that of photometric surveys, we also recorded which of the simulated systems would present eclipses by checking whether the standard eclipse condition 
\begin{equation}
    \cos i  < \frac{R_1+R_2}{a}\label{eq:eclipses}
\end{equation}
was fulfilled,
where $R$ are the stellar radii and $a=a_1+a_2$ is the semi-major axis of the relative orbit. In doing so, we adopt an average mass-radius relation \cite{eker2018}
\begin{equation}
\frac{R}{R_\odot}=\left(\frac{M}{M_\odot}\right)^{0.72}
\end{equation}
and we ignore eccentricity effects (but most eclipsing binaries have low eccentricity given they are restricted to short periods). We also ignore the fact that non-eclipsing short-period systems can be detected through ellipsoidal variations or mutual-illumination effects, and we did not include any further detection criteria (e.g., depth of the eclipses, phase coverage, photometric noise, ...). While our simulations are simplified, the results are clearly cut. The probability to display eclipses drops significantly for systems with orbital periods more than a few days (Extended Data Figure~2), %\ref{f:detect}),
which prevents photometric surveys to adequately map the range of periods relevant for binary evolution (orbital periods of up to several years).  

 The \bloem\ detection probability is well above 0.9 for orbital periods up to 3 months but drops rapidly after that (see Extended Data Figures~2 and 3).%\ref{f:detect} and \ref{f:detect_2d}).
 Integrated over a period range of up to one year, the overall detection rate is $0.89\pm0.03$. Integrated over the full period range considered ($\log_{10}\left(P/\mathrm{d}\right) < 3.5$), our detection rate is $0.68\pm0.06$, where the $1\sigma$ uncertainties are computed from the dispersion observed in the results of 10\,000 simulations and include uncertainties in the parent orbital distributions (Eqs.~\ref{eq:Pidx}--\ref{eq:Eidx}), the measurements, the sample size, and the random orientation of the orbital plane of the binaries in 3-dimensional space.

% Applied to the results of the previous section, this yields a bias-corrected binary fraction of $0.68^{+0.08}_{-0.06}$, which is close to the bias-corrected binary fractions estimated for Galactic clusters ($0.69\pm0.09$ \cite{sana2012}), the Cyg OB2 region ($\approx 0.55$, \cite{KK2014}) and the LMC Tarantula region ($0.51\pm0.03$, \cite{sana2012}, later revised to $\approx 0.58$ \cite{tmbm1,Sana2017}) using similar methods and similar types of surveys.

The bias correction methods used in \cite{sana2012,KK2014,vfts8} and in the present work so far only consider the RV signal of the most luminous star of a putative binary, but ignore the impact of the companion on the accuracy of the RV measurements. Indeed, contamination of the primary line profile by that of the secondary star tends to decrease the amplitude of the RV signal, making it harder to detect a pair as a binary system \cite{sana2011}. This is especially true at large orbital separations and \mbox{(near-)equal} line intensity ratios \cite{Bodensteiner2021, Banyard2022} (see also Extended Data Figures~2 and 3.)%\ref{f:detect} and \ref{f:detect_2d}.) 
The effect is worsened for broader spectral lines, independently of whether the broadening is induced by the observational technique (i.e., lower resolution spectroscopy) or by physical processes (e.g., rotational broadening). The effect of this line-blending bias is also more important in binaries with a lower primary mass  ($M_1$) as the amplitude of the RV signal scales with $M_1^{1/3}$ for a fixed mass ratio and orbital period.  We used specific calibrations of the line-blending bias developed following the strategy outlined in \cite{Bodensteiner2021, Banyard2022} and adapted to O-type stars observed with the low-resolution mode of the FLAMES/Giraffe instrument used in \bloem. We assumed a constant projected equatorial spin of 100~\kms.   Extended Data Figures~2 and 3%\ref{f:detect} and \ref{f:detect_2d}  
compare the \bloem\ binary detection rate with and without accounting for the line-blending bias and shows that binaries with periods close to one year and (near-)equal mass-ratio systems are the most affected. Under these assumptions, the overall detection rate of the survey decreases to $0.58\pm0.06$, yielding an intrinsic binary fraction of $\mathcal{F}_\mathrm{bin}=0.77\pm{0.08}$.

Larger rotation rates further increase the importance of the line-blending bias. Detailed simulations including the individual rotation rates of \bloem\ stars are, however, beyond the scope of this work. Applying a similar correction including the line-blending bias to earlier OB-type spectroscopic surveys in the Milky Way and the LMC would also be interesting but is beyond the scope of the present work. Yet, our computations show that the multiplicity fractions derived while ignoring the effect of line blending, such as those quoted in the main text, are most likely underestimated, making our conclusions on the importance of massive binaries even more robust.

\subsection*{Intrinsic binary fraction and orbital period distribution}

Obtaining individual correct orbital solutions with only nine epochs for the majority of the detected binaries is unrealistic, especially given the short time base and numerous aliases induced by the sparse and irregular sampling of the observations. However, we can still obtain important statistical constraints, e.g.\ on the orbital period distribution. In particular, \cite{sana2012} showed that the distribution of shortest time lapses $\delta t$ in which significant ($\Delta RV/\sigma_{\Delta RV}> 4\sigma$), large $(\Delta RV > 20 \rm{km\,s}^{-1})$ RV variations occur is a sensitive probe of the orbital period distribution. 

To illustrate the sensitivity of shortest time lapses to the orbital period distribution, we compute the distributions of $\delta t$ for simulated observational campaigns and repeat the process with different assumptions on the orbital period index $\pi$. Extended Data Figure~4%\ref{f:dHJD} 
illustrates the locus of simulated distributions and compares it to the \bloem\ observed distribution. It reveals that adopting an intrinsic log-period distribution given by power-law with an index $\pi\approx 0$ provides a reasonable representation of the observed $\delta t$ distribution. %Figure compares the distribution of `agreement' probabilities computed from a 2-sided Kuiper test of hypothesis between the \bloem\ observations and the simulated distributions. 

We use the Kuiper statistics $D$ as a goodness-of-fit criteria to evaluate how well simulated and observed distributions compared \cite{Arsham1988}.  The Kuiper statistic $D$ is defined as $D=D^+ + D_-$, where $D^+$ and $D_-$ are, respectively, the maximum positive and maximum negative deviations between the empirical cumulative distribution function (CDF) if the observational sample and that of a known distribution (one-sided Kuiper test) or another sampled distribution (two-sided Kuiper test). Unlike the Kolmogorov-Smirnov (KS) test, the Kuiper test accounts for deviations in both directions and is sensitive to differences in the tails and the center of the distributions. Because we compare our observed distribution with simulated distributions of an equally large sample, we use a two-sided Kuiper's test of hypothesis. The latter does not allow us to reject \"Opik's law at the 0.1-significance level in 99\%\ of the 10\,000 simulated distributions (compared to 55\%\ when adopting $\pi=-0.5$), again suggesting that \"Opik's law adequately captures the time variability of stars in our sample.

As a final step, we build a grid of synthetic populations of O-type stars, varying the intrinsic binary fraction $\mathcal{F}_\mathrm{bin}$ from 0.50 to 0.90 in steps of 0.02, and orbital period index $\pi$ from $-0.80$ to $+0.50$ in steps of 0.05. We keep the indexes of the mass-ratio and eccentricity distributions constant at $\kappa=0$ and $\eta=-0.5$, in agreement with constraints obtained for Milky Way and LMC O-type binaries \cite{sana2012,tmbm1,tmbm6}. As above, we apply the temporal sampling, RV uncertainties and detection criteria of the \bloem\ survey.  Each simulated population contains 1.39 million objects so that the statistical uncertainties of the simulated results are two orders of magnitude smaller than those due to the \bloem\ sample size. We then compute the binary yield and the $\delta t$ distribution of each mock observing campaign and we compare this with observed values. Specifically, we compute the probability $\mathcal{P}\left({N}_\mathrm{bin}\right)$ to detect $N_\mathrm{bin}=62$ binaries of a sample of $N=139$ given the simulated binary yield $f_\mathrm{sim}$ and binomial statistics. The log-likelihood map that we obtain is displayed in Extended Data Figure~5,%\ref{f:proba1}, 
where $\mathcal{L}=-\log_{10} \mathcal{P}\left({N}_\mathrm{bin}\right)$. The best estimate of the intrinsic binary fraction $\mathcal{F}_\mathrm{bin}$ depends on the period distribution index $\pi$. In all cases however, the binary fraction consistently exceeds 0.5 across the range explored, so that we can robustly conclude that a majority of objects are binaries.

Restricting ourselves to those objects that satisfy our binary detection criteria, we also compare the predicted distribution of the smallest time difference between any pair of RVs that satisfy the detection criteria and compare these to the observed distribution using the value of the  two-sided Kuiper statistics ($P_\mathrm{Kuiper}$, e.g.\cite{Press92}). The $P_\mathrm{Kuiper}$-probability map obtained as a function of $\mathcal{F}_\mathrm{bin}$ and $\pi$ is displayed in Extended Data Figure~5.%\ref{f:proba1}.
It shows that the index of the period distribution is mostly independent from the value of the intrinsic binary fraction. The best representation is obtained with $\pi=+0.10$, and $\pi<-0.24$ and $\pi>+0.40$ are rejected at the 0.1-significance level.

To combine these two sets of constraints, we follow the approach outlined and validated in \cite{vfts8} and adopt a metric based on the products of the binomial and Kuiper probabilities. The advantage of this approach resides in its simplicity, where regions of low probability are given a much lower metric than regions where both probability values are high.  The obtained 2D-peak distribution (Figure~\ref{f:proba_combined}) is not a measure of the probability of realization of such a pair so that one cannot simply integrate the highest-confidence intervals to obtain uncertainties. Instead, \cite{vfts8} shows that the full width at half maximum of the peak provides a good approximation of the $1\sigma$ uncertainties on the best-fit parameter and we follow this approach here. We thus obtained best-fit estimates $\mathcal{F}_\mathrm{bin}=0.70^{+0.11}_{-0.06}$ and $\pi=+0.10^{+0.20}_{-0.15}$.

\subsection*{No evidence for a trend with metallicity}

\subsubsection*{Intrinsic binary fraction}
The  intrinsic O-star binary fraction that we obtain  for the SMC ($\mathcal{F}_\mathrm{bin}(Z_\odot/5)=0.70^{+0.11}_{-0.06}$) exceeds that at Solar ($\mathcal{F}_\mathrm{bin}(Z_\odot)=0.69\pm0.09$, \cite{sana2012}), and half-Solar metallicity ($\mathcal{F}_\mathrm{bin}(Z_\odot/2)=0.58\pm0.04$, \cite{tmbm1}). To quantitatively investigate the presence of a trend with metallicity, as seen in lower-mass stars \cite{Moe2019,Mazzola2020}, we perform a linear regression of the intrinsic binary fraction as a function of metallicity following $\mathcal{F}_\mathrm{bin}(Z) = a+b*\log_{10} (Z/Z_\odot)$.  %The resulting 2-d posterior on the intercept and slope, as well as the marginalized 1-d posteriors, are displayed in Figure~\ref{fig: posterior}. 
The best-fit intercept at $Z=Z_\odot$ is $a=(0.58 \pm 0.06)/\mathrm{dex}$ and the slope of binary fraction vs.~log-metallicity is $b=(-0.11 \pm 0.15)/\mathrm{dex}$ (Figure~\ref{f:reg}).
This slope is $0.5\sigma$ from, hence consistent with, zero within our measurement uncertainties so provides no support for a relation between metallicity and intrinsic binary fraction. However, our results do not allow us to reject a slight metallicity effect. Indeed a slope of $(-0.16\pm0.01)/\mathrm{dex}$, as obtained in solar-mass data from \cite{Moe2019}, remains within $1\sigma$ (see comparison in Figure~\ref{f:reg}).

\subsubsection*{Orbital period distribution}

The index $\pi$ of the period distribution was derived from two Milky Way samples to be $\pi = -0.55\pm 0.22$ \cite{sana2012} and $\pi \approx -0.22$ \cite{KK2014} using orbital period ranges of 1.4 to 3200 days, and 1.4 to 2000 days, respectively. Similar estimates from a sample of 354 O stars in the LMC Tarantula region yielded $\pi=-0.45\pm0.30$, using an orbital period range of 1.4 to 3200 days. The discovery of orbital periods shorter than 1.4~days led \cite{tmbm1} to modify the lower limit of the period distribution for the LMC Tarantula sample to 1~day. As discussed in \cite{tmbm1}, the chosen upper and lower limits of the period distribution impact the best fit power-law index. Consequently, \cite{tmbm1} revised the LMC results of \cite{vfts1} using the new lower boundaries and obtained $\pi \approx-0.1$, increasing the intrinsic binary fraction from $\mathcal{F}_\mathrm{bin}=0.51\pm0.04$ to $\approx 0.58$. A similar revision of the Milky Way results has not been performed yet, but changing the lower limit of the period distribution fit will inevitably flatten the best-fit period distribution so that, overall, there is no significant evidence supporting an effect of metallicity on the shape of the orbital period distribution of O stars. This matches the results of other studies that compare the (measured) orbital periods of O- and B-type stars in low- and high-density star forming region, and in the LMC and SMC and show no significant differences either \cite{own,BBC1,Banyard2022}, suggesting that the distribution of orbital period is universal across a wide range of environments and metallicities.

While small differences are possibly observed in the intrinsic binary fraction between the LMC Tarantula region on the one hand, and the Milky Way and SMC open clusters and field regions on the other, our results indicate that the \bloem\ measurements are also statistically compatible with {\"O}pik's law. So far there is thus no significant observational evidence for a metallicity effect on either the binary fraction or orbital period distributions of O-type stars, and that massive binaries close enough to interact during their lifetime are ubiquitous in the metallicity range $0.2\dots1 Z_\odot$.

\subsection*{Binary interaction rate}
The rate of binary interaction can be estimated by direct integration of the orbital period distribution up to a critical period beyond which systems are believed to be too wide to interact. The type of stars involved (O-type, B-type, lower mass) and the outcome of the interaction (stripped stars, mass gainers, mergers) additionally require knowledge of the mass-ratio distribution. 
The orbital-parameter distributions can be modified by evolutionary effects. For example, mass loss through stellar winds widens orbits, while binary interaction flattens the period distributions because the closest systems interact first and many merge. Given the weak winds of O-type stars at SMC metallicity, and the fact that our sample is formed by main-sequence stars, we argue that these effects have a limited impact on the observed distributions and that the observed distributions could be close to pristine.

Aiming at offering a direct comparison with earlier results obtained at higher metallicity, we adopt the same hypotheses and critical periods. Specifically, we adopt a flat mass-ratio distribution between 0.1 and 1.0 (as derived from Milky Way and LMC samples \cite{sana2012,tmbm6}), and conservatively assume that all systems with orbital periods shorter than 1500 days interact, consistent with \cite{sana2012}. Similarly, we assume that all systems with an orbital period shorter than 6 days interact before either of the binary components leaves the main sequence.  We then count all O-type stars that interact, irrespective of whether they are a primary or secondary, and compare this to the total number of O-type stars. We estimate that $76\pm8$\%\ of SMC stars that are born as O~stars are also in a binary with an orbital period shorter than 10$^{3.5}~$d$~\approx8.7$~yr. As a result,
$68\pm7$\%\ of all stars born as O~type interact, at least $18\pm5$\%\ of which do so before leaving the main sequence (hence $12\pm4$\%\ of all stars born as O stars). 

The interaction limits adopted here and in \cite{sana2012} are quite conservative. In practice, systems with orbital periods of up to 10 years probably interact and even longer-period systems interact if they are in sufficiently eccentric binaries. For example, \cite{Moe2017} consider interaction through Roche-lobe overflow at orbital periods up to nearly 4000 days, and even wider systems still interact through wind mass transfer \citep[e.g.][]{Ercolino2024}. Similarly, the $P=6$~day limit for main-sequence interaction is also conservative. How large massive stars become during core hydrogen burning is not well known \citep[e.g.][]{Castro2018} and substantially wider systems may still interact while both stars are on the main sequence \citep[][e.g.]{Sen2022}. Finally, the range of initial orbital periods in which binaries interact depends on metallicity. At low metallicity, stars are generally more compact during their main-sequence evolution, but they may be larger in later stages as reduced stellar winds allow for larger stellar-envelope expansion after the main sequence. Overall, we expect this to further enhance the fraction of stars that interact, especially at high mass.

%%%%%%%%%%%%%%%%%%%%%%%%%%%%%%%%%%%%%%%%%%%%%%%%%%%%%%%%%%%%
%%%%%%%%%%%%%%%%%%%%%%%%% METHODS %%%%%%%%%%%%%%%%%%%%%%%%%%
%%%%%%%%%%%%%%%%%%%%%%%%%%%%%%%%%%%%%%%%%%%%%%%%%%%%%%%%%%%%

\subsection*{Data availability} \label{s:data_avail}
The raw data used are publicly available in the ESO archive ({\tt https://www.eso.org/archive}). The normalized spectra will be made available on the ESO Phase 3 webpage (\url{https://www.eso.org/sci/observing/phase3.html}) upon completion of the program. Both databases can be queried using the \bloem\ ESO program ID 112.25R7 for observations taken in 2023. Tables 1 and 2 are available from the Centre de Donn\'ees astrophysiques de Strasbourg (CDS) via anonymous ftp to cdsarc.u-strasbg.fr (130.79.128.5) or
via \url{ https://cdsarc.cds.unistra.fr/viz-bin/cat/J/other/NatAs/}Vol.Num

\subsection*{Code availability} \label{s:code_avail}
The RV measurement cross-correlation code is available on GitHub ({\tt https://github.com/TomerShenar}).

\section*{Acknowledgements} \label{s:ack}
Based on data collected at the European Southern Observatory (ESO) under program ID 112.25R7. The research leading to these results has received funding from the European Research Council (ERC) under the European Union's Horizon 2020 and Horizon Europe research and innovation programme (grant agreement numbers 772225: MULTIPLES, 772086: ASSESS and 945806: TEL-STARS, ADG101054731: Stellar-BHs-SDSS-V, and 101164755: METAL). This research was supported by the Israel Science Foundation (ISF) under grant number 0603225041. The authors acknowledge support from the Science and Technology Facilities Council (research grant ST/V000853/1 and ST/V000233/1), UK Research and Innovation (UKRI) and the UK government's ERC Horizon Europe funding guarantee (grant number: EP/Y031059/1), a Royal Society University Research Fellowship (grant number: URF{\textbackslash}R1{\textbackslash}231631), a Royal Society--Science Foundation Ireland University Research Fellowship, the German Deutsche Forschungsgemeinschaft (Project-ID 496854903,  445674056, and 443790621,  Germany's Excellence Strategy EXC 2181/1-390900948), the Klaus Tschira Foundation, the JSPS Kakenhi Grant-in-Aid for Scientific Research (23K19071) and international fellowships (at the Graduate school of Science, Tokyo University), the Australian Research Council (ARC) Centre of Excellence for Gravitational Wave Discovery (OzGrav; project number CE230100016), the Deutsches Zentrum f\"ur Luft und Raumfahrt (DLR) grants FKZ 50OR2005 and 50OR2306, Agencia Espa\~nola de Investigaci\'on (AEI) of the Spanish Ministerio de Ciencia Innovaci\'on y Universidades (MICIU) and the European Regional Development Fund, FEDER and Severo Ochoa Programme (grants PID2021-122397NB-C21 and CEX2019-000920-S), the  NextGeneration EU/PRTR and MIU (UNI/551/2021) trough grant Margarita Salas-UL,  the CAPES-Br and FAPERJ/DSC-10 (SEI-260003/001630/2023), MCIN/AEI/10.13039/501100011033 by "ERDF A way of making Europe" (grants PID2019-105552RB-C41 and PID2022-137779OB-C41, PID2021-125485NB-C22,  CEX2019-000918-M) funded by MCIN/AEI/10.13039/501100011033 (State Agency for Research of the Spanish Ministry of Science and Innovation) and SGR-2021-01069 (AGAUR), the Spanish Government Ministerio de Ciencia e Innovaci\'on and Agencia Estatal de Investigaci\'on (10.13\,039/501\,100\,011\,033;  grant PID2022-136\,640~NB-C22), the Consejo Superior de Investigaciones Cient\'ificas (CSIC;  grant 2022-AEP~005), the Polish National Agency for Academic Exchange (BEKKER fellowship BPN/BEK/2022/1/00106) and National Science Center (NCN, Poland;  grant number OPUS 2021/41/B/ST9/00757), the ``La Caixa'' Foundation (ID 100010434) under the fellowship code LCF/BQ/PI23/11970035, the Research foundation Flanders (FWO) PhD fellowship under project 11E1721N and senior postdoctoral fellowship under number 12ZY523N, and the Netherlands Research Council NWO (VIDI 203.061 grant).

\section*{Author Contributions Statement} \label{s:contrib}
JB, TS, HS prepared and obtained the observations, and performed data reduction. Together with NB, DL,  LM, LP, and JV, they performed the scientific validations of the data and discussed and implemented the strategy for multiplicity analysis. JB and PC performed spectral typing. TS measured the radial velocities. HS performed the bias correction and multiplicity analysis. IM and TS performed the regression of the binary fraction with metallicity. JB, TS, HS prepared the manuscript draft with input from NL and SdM. All co-authors participated in writing the observational proposal, observational strategy, sample selection, and were given the opportunity to participate in the discussion of the scientific results, and to comment on the manuscript.

\section*{Competing Interests Statement}\label{s:declare}
Authors have no competing interests.

\newpage
\section*{Main Figures}\label{s:fig_captions}

%-------------------------- Figure 1 -----------------------------
\renewcommand{\figurename}{Figure}
%-------------------------- End Figure 1 ------------

%-------------------------- Figure 2 --------------------------
%-------------------------- End Figure 2 ------------

%-------------------------- Figure 3 -----------------------------
%-------------------------- End Figure 3 ------------

%-------------------------- Figure 4 ---------------------------
%-------------------------- END Figure 4 ------------

%\section*{Supplementary Information}
%Supplementary Information is available for this paper.

%\subsubsection*{Reprints and permissions information is available at \url{www.nature.com/reprints.}}

\clearpage
%\normalem
\bibliographystyle{sn-nature}
%\bibliography{bloem2-bibtex}

\section*{Extended Data Figure}\label{s:extended_data}

\renewcommand{\figurename}{Extended Data Figure}
%\counterwithin{figure}{section}
\setcounter{figure}{0}  

%--------------------------  Extended Data Figure 1 -----------------------------
\begin{figure*}[h!]
\centering
\includegraphics[width=0.99\linewidth]{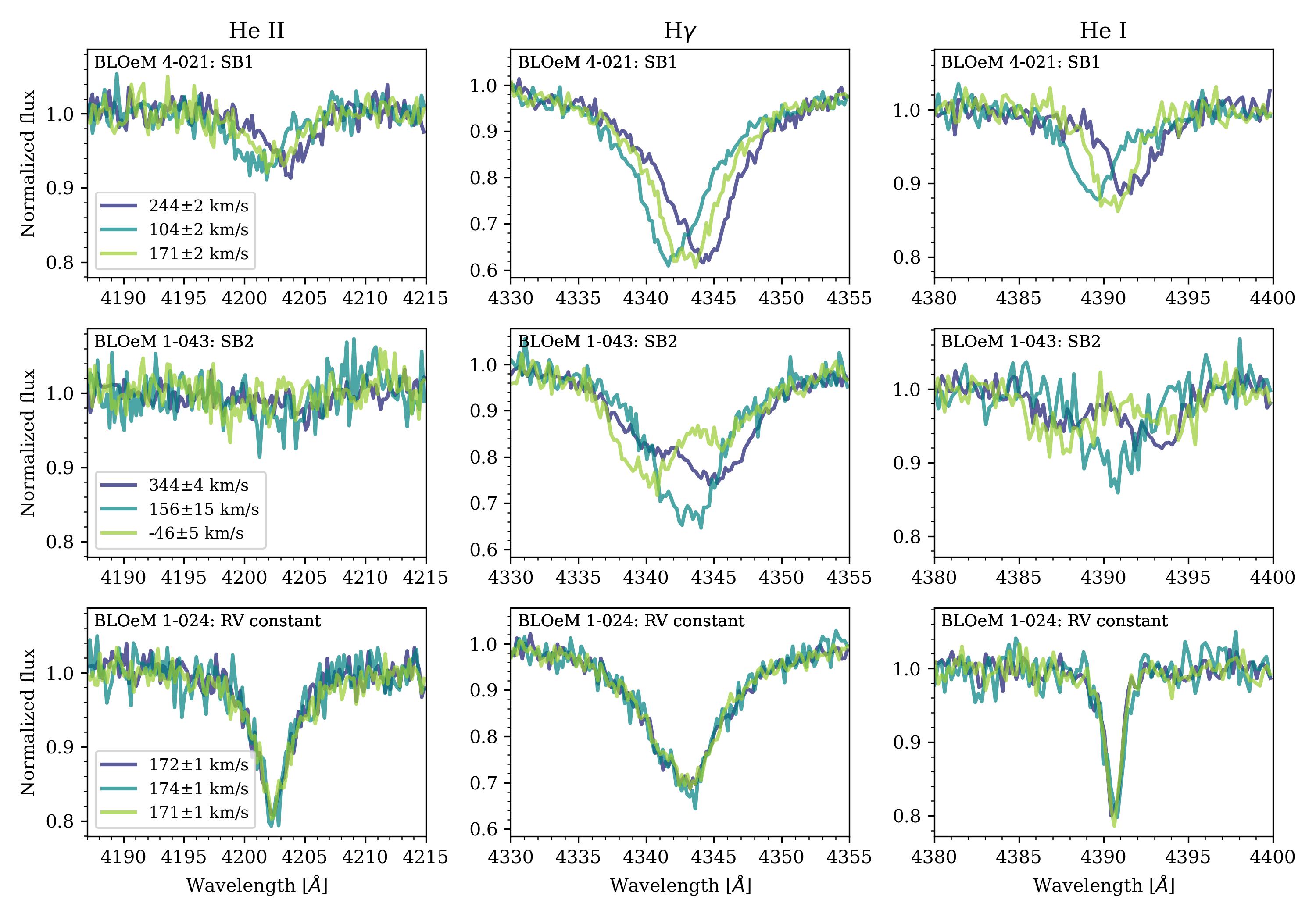}
\caption{\textbf{\textbar\ Examples of VLT FLAMES/Giraffe spectra of three \bloem\ targets.} The top row shows a single-lined spectroscopic binary (SB1); the middle row, a double-lined spectroscopic binary (SB2); and the bottom row, a source with no statistically significant radial-velocity (RV) variations. From left to right, each column displays a different spectral range, centered on lines of He{\sc ii}~\l4200, H$\gamma$, and He{\sc i}~\l4390, respectively. Three different observing epochs are provided for each target, with the measured RVs and their standard deviations listed in the left column.}
\label{f:spectra}
\end{figure*}
%-------------------------- END Extended Data Figure 1 ------------

%--------------------------  Extended Data Figure 2 -----------------------------
\begin{figure*}
\centering
\includegraphics[width=0.99\linewidth]{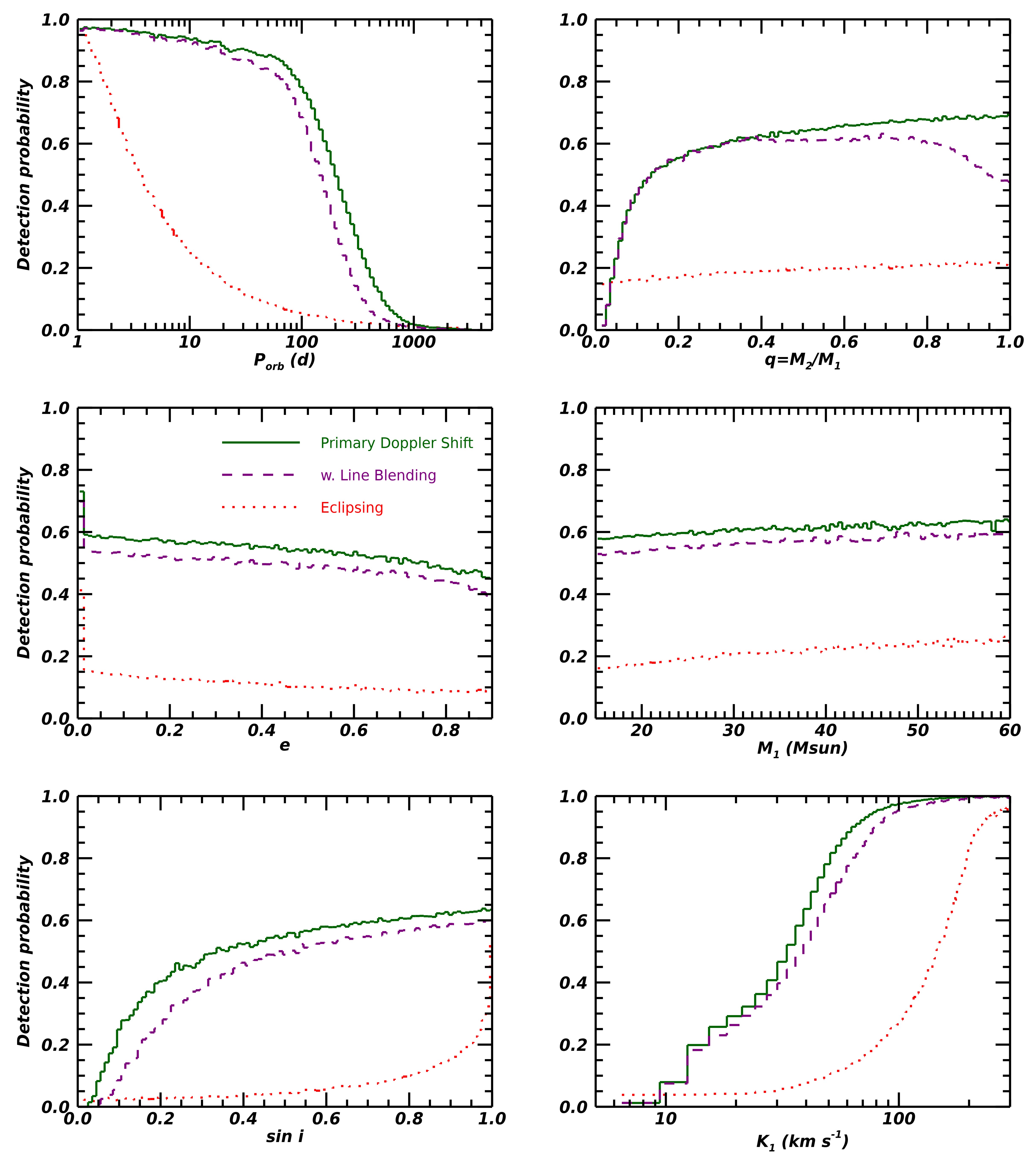}
\caption{\textbf{\textbar\ O-type binary detection probability curves of the BLOeM survey.} Different panels show detection probabilities projected on different orbital properties. The plain green lines are obtained through the standard approach used in the main part of the paper, which rely on the amplitude of the Doppler shift of the primary star. The dashed, purple lines also include the line-blending detection bias, which reduces detection at longer periods, and (near-)equal mass ratio. The dotted, red line indicates the fraction of the simulated systems that display eclipses. The upper left panel is identical to Figure~\ref{f:detect_p}.}
\label{f:detect}
\end{figure*}
%-------------------------- END Extended Data Figure 2 ------------

%-------------------------- Extended Data Figure 3 -----------------------------
\begin{figure*}
\centering
\includegraphics[width=0.99\linewidth]{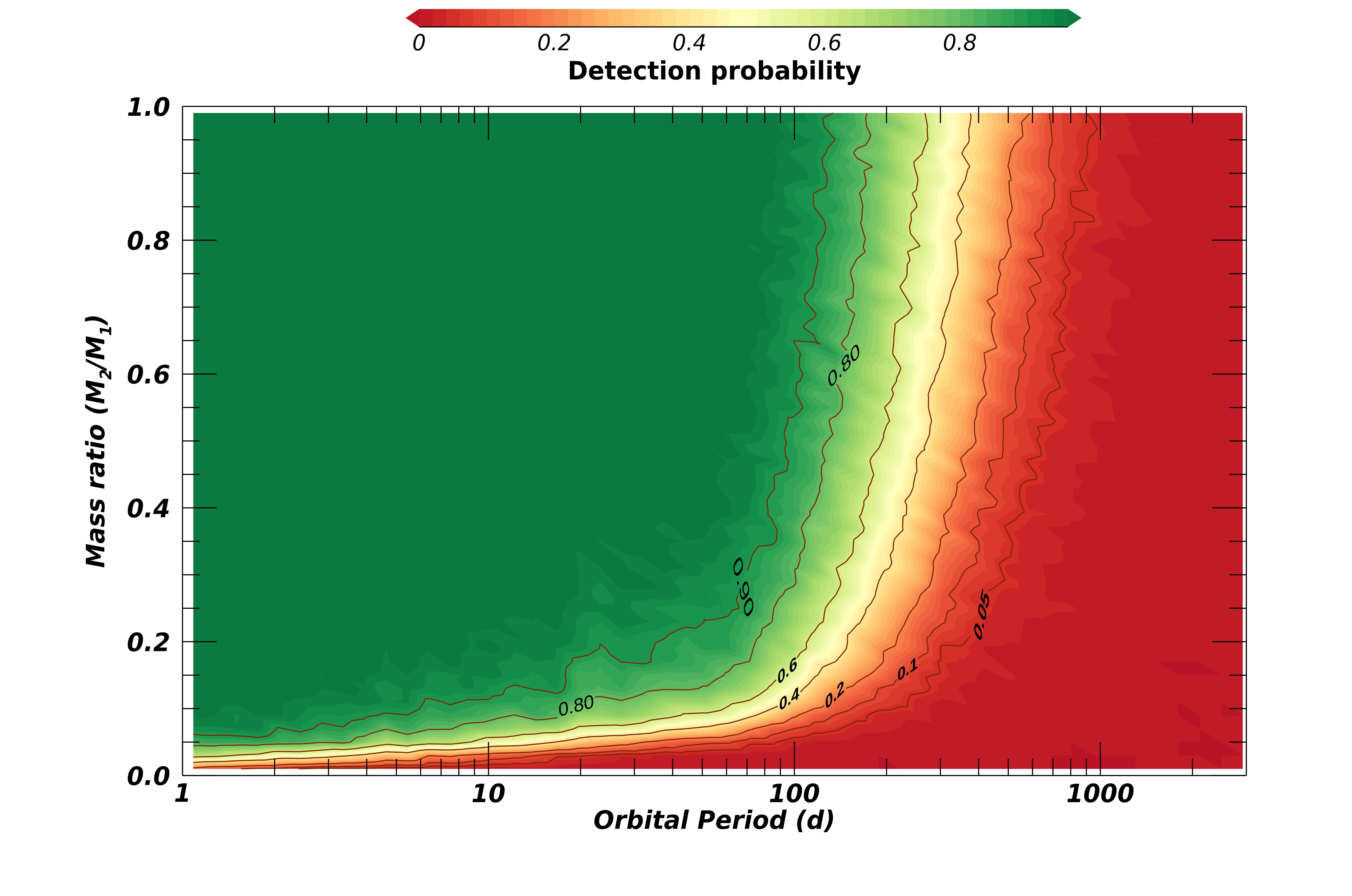}
\includegraphics[width=0.99\linewidth]{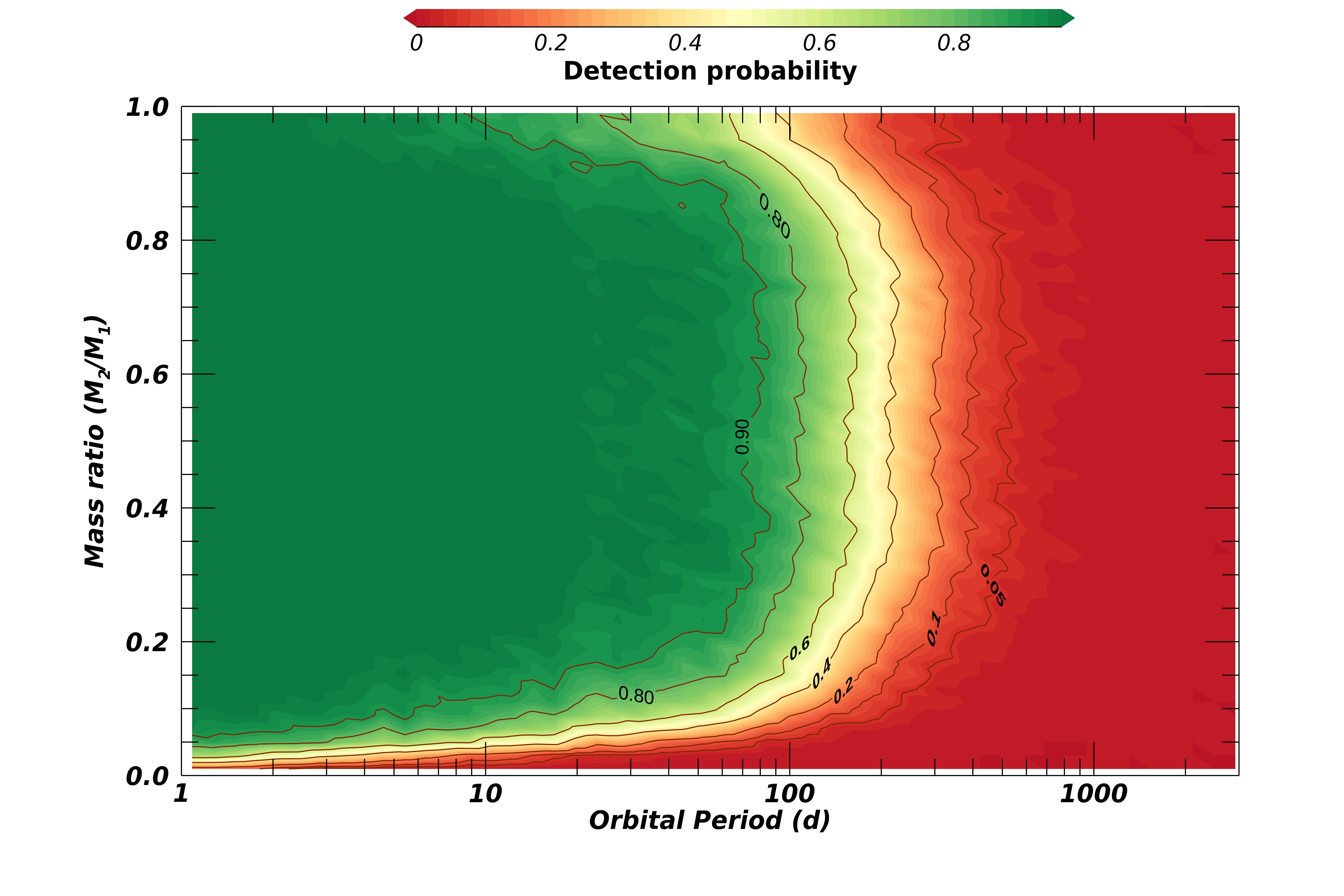}
\caption{\textbf{O-type binary detection probability maps of the \bloem\  survey.} The panels displayed the binary detection probabilities projected on the mass-ratio vs. orbital period plane. The top (resp. bottom) panel ignores (resp. includes) the line-blending bias. These figures show that the BLOeM survey has excellent detection capability for periods shorter than about 3 months, and that the sensitivity drops quickly for periods longer than 5 or 6 months.}
\label{f:detect_2d}
\end{figure*}
%-------------------------- END Extended Data Figure 3 ------------

%-------------------------- Extended Data Figure 4 -----------------------------
\begin{figure*}
\centering
\includegraphics[width=0.9\linewidth]{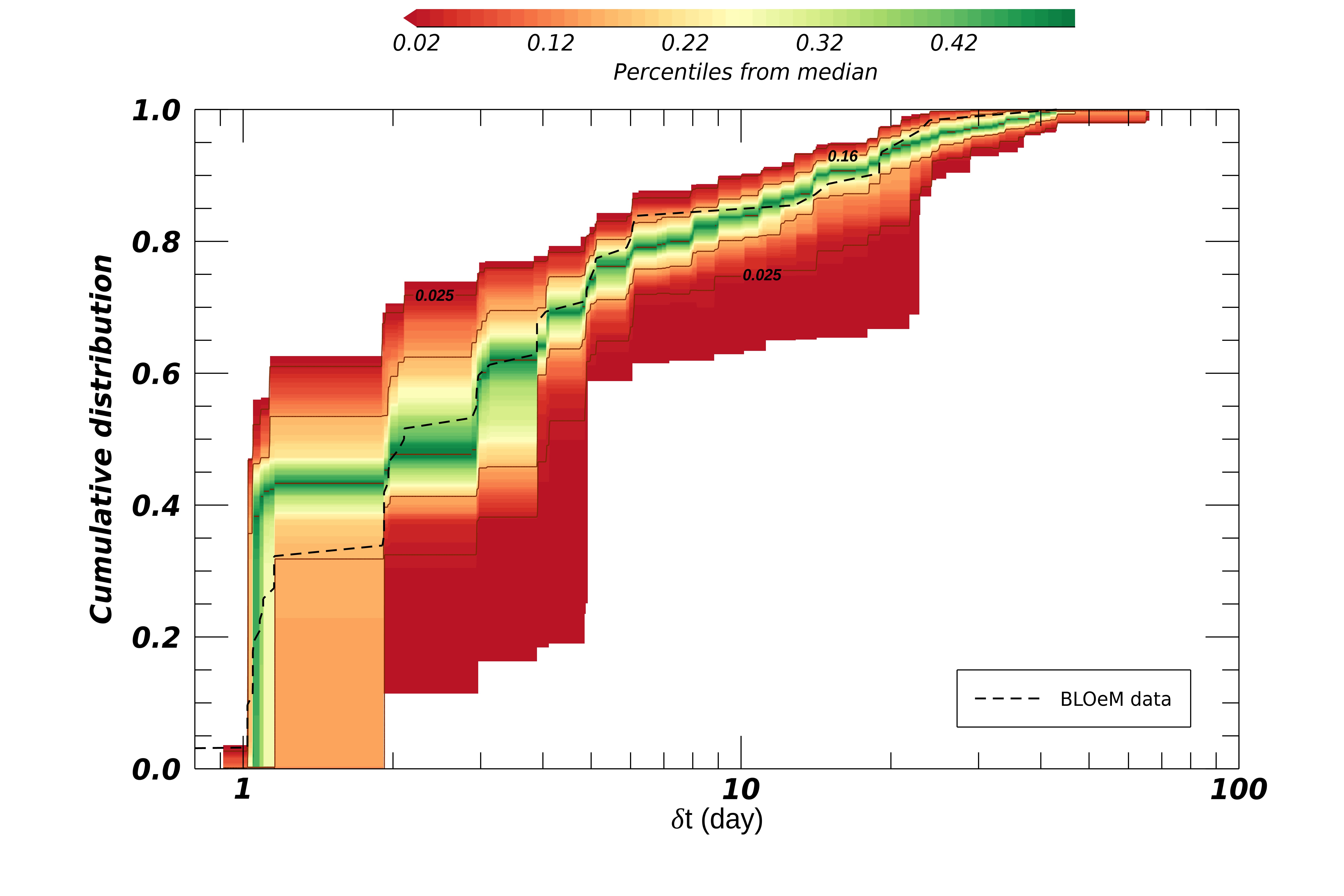}
\includegraphics[width=0.9\linewidth]{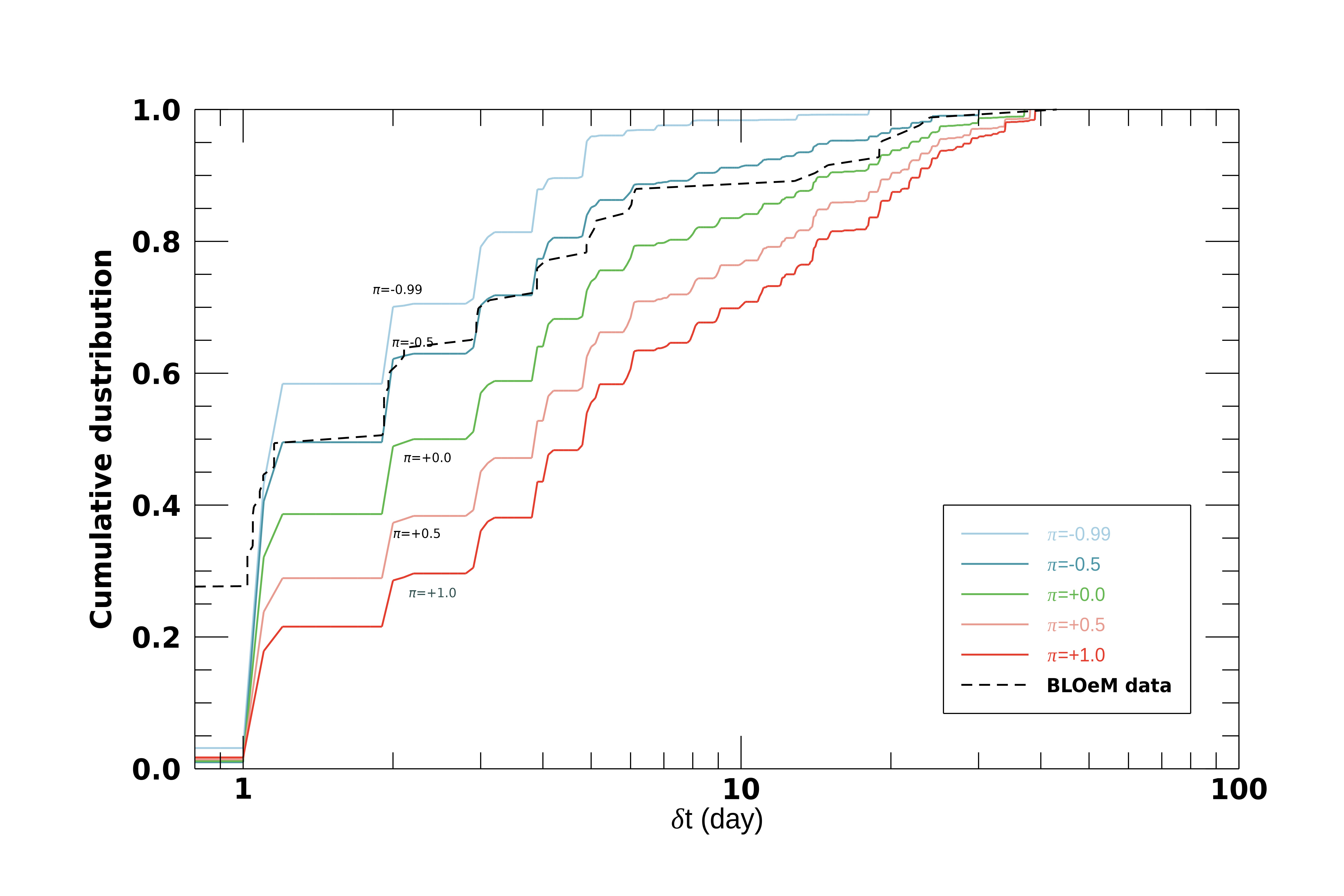}
\caption{\textbf{Cumulative distribution of smallest time differences} $\delta t$ between observations that exhibit significant ($\Delta RV > 4 \sigma_{\Delta RV}$) and large $(\Delta RV > 20 \rm{km\,s}^{-1})$ radial velocity variations in the \bloem\ O-star sample (dashed line) and for various mock populations. Top panel: observed \bloem\ distribution overlaid on the confidence areas covered by simulated distributions produced with orbital period, mass-ratio and eccentricity distributions listed in Eqs.~\ref{eq:Pidx} to \ref{eq:Midx} (i.e., with $\pi=0.0\pm0.2$, $\kappa=0.0\pm0.2$ and $\eta=-0.5\pm0.2$). Bottom panel: similar to the top panel but varying the power law index $\pi$ of the orbital period distribution ($f_{\log P}\propto(\log P)^\pi$). Only the median distribution is provided for clarity. See Methods section for further information.}
\label{f:dHJD}
\end{figure*}
%--------------------------  END Extended Data Figure 4 -----------------------------

%-------------------------- Extended Data Figure 5 -----------------------------

\begin{figure*}
\centering
\includegraphics[width=0.96\linewidth]{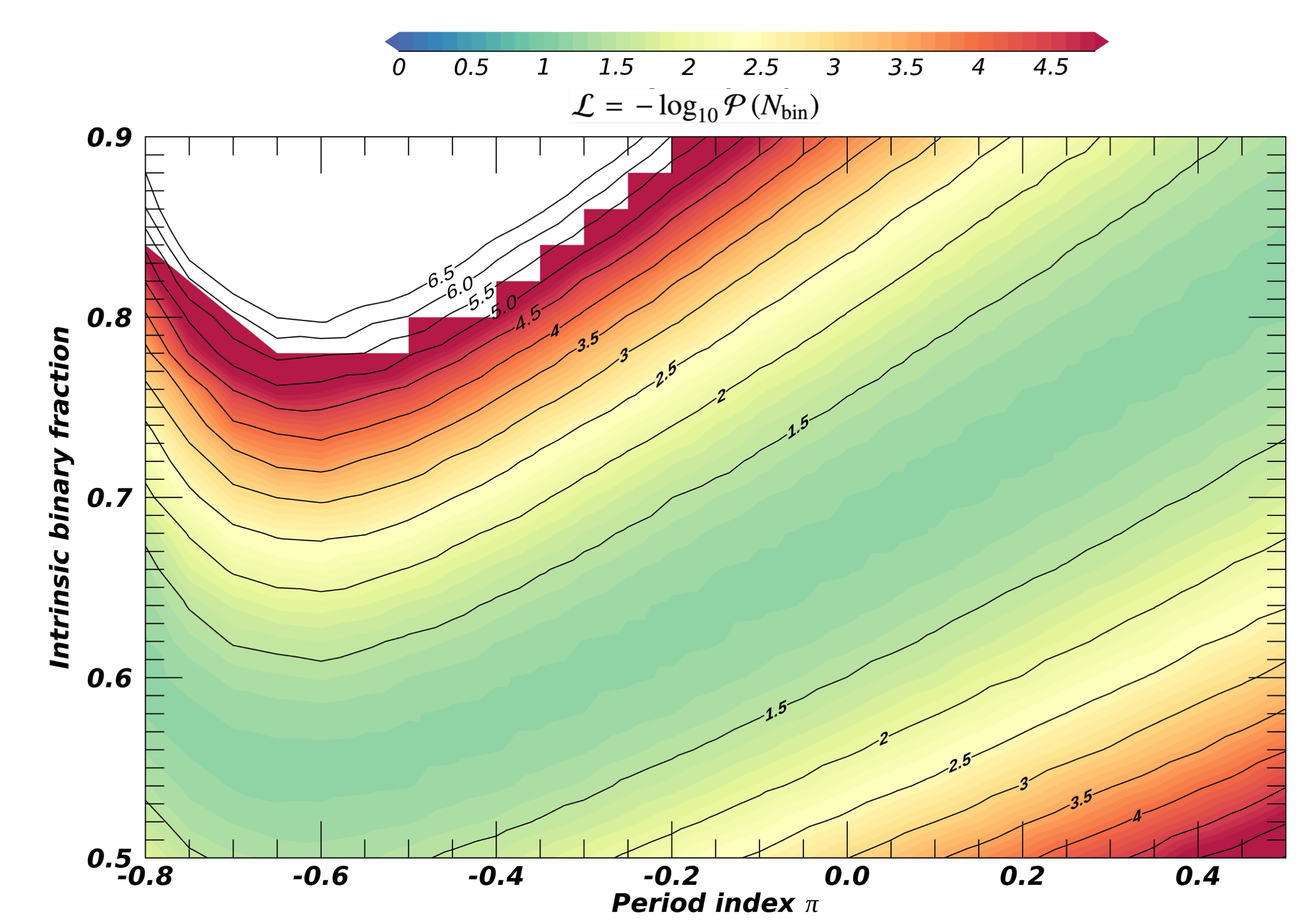}
\vspace*{0.5cm}
\includegraphics[width=0.99\linewidth]{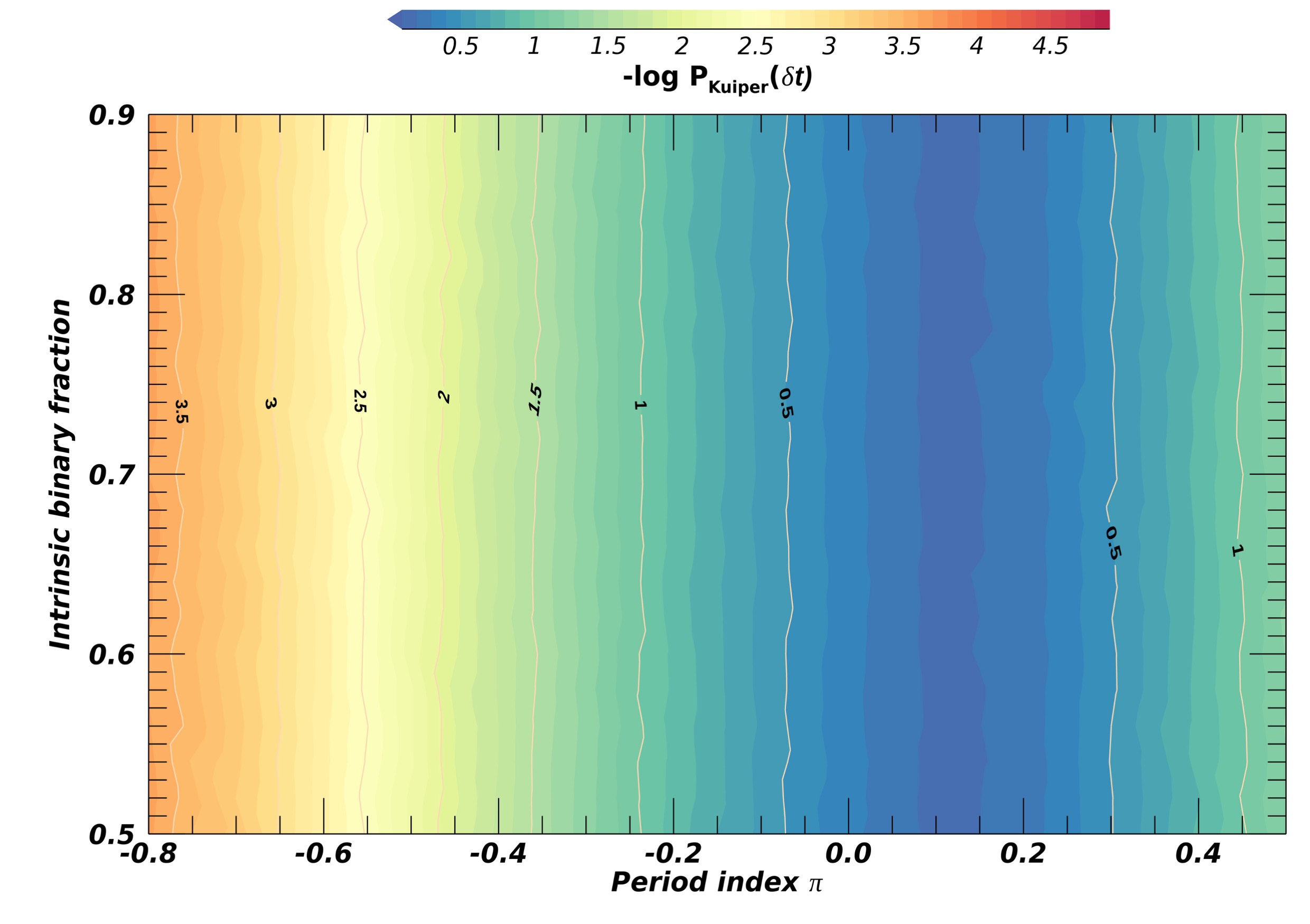}
\caption{\textbf{Probability maps} for simulations to reproduce the observed number of binaries (top panel) and the distribution of shortest time lapses for a system to meet our binary criteria (bottom panel).}
\label{f:proba1}
\end{figure*}

\end{document}